\documentclass[twocolumn]{aastex631}

\usepackage{amsmath}
\usepackage{amssymb}
\usepackage{natbib}
\usepackage{ulem}
\hypersetup{linkcolor=red,citecolor=blue,filecolor=cyan,urlcolor=magenta}
\usepackage{bbding}
\usepackage{graphicx}
\usepackage{cases}
\usepackage{color, soul}
\usepackage{array,multirow}

\newcommand{\MO}{$M_\odot$}

\newcommand{\Chandra}{\textit{Chandra}}

\newcommand{\rxcj}{RXCJ1504.1-0248}

\shorttitle{The relation between the cool-core radius and the host galaxy clusters}
\shortauthors{Ng \& Ueda}
\graphicspath{{../}{figs/}}
\begin{document}

\title{
The relation between the cool-core radius and the host galaxy clusters: thermodynamic properties and cluster mass
}

\correspondingauthor{Shutaro Ueda}
\email{sueda@asiaa.sinica.edu.tw}

\author[0000-0003-4195-6300]{FanLam Ng}
\affiliation{Academia Sinica Institute of Astronomy and Astrophysics (ASIAA), No. 1, Section 4, Roosevelt Road, Taipei 10617, Taiwan}
\affiliation{Yuanpei College, Peking University, Yi He Yuan Road 5, Hai Dian District, Beijing 100871, China}

\author[0000-0001-6252-7922]{Shutaro Ueda}
\affiliation{Academia Sinica Institute of Astronomy and Astrophysics (ASIAA), No. 1, Section 4, Roosevelt Road, Taipei 10617, Taiwan}
\affiliation{Institute of Astronomy, National Tsing Hua University, Hsinchu 30013, Taiwan}

%\maketitle


\begin{abstract}
We present a detailed study of cool-core systems in a sample of four galaxy clusters (RXCJ1504.1-0248, A3112, A4059, and A478) using archival X-ray data from the \Chandra ~X-ray Observatory. Cool cores are frequently observed at the centers of galaxy clusters and are considered to be formed by radiative cooling of the intracluster medium (ICM). Cool cores are characterized by a significant drop in the ICM temperature toward the cluster center. We extract and analyze X-ray spectra of the ICM to measure the radial profiles of the ICM thermodynamic properties including temperature, density, pressure, entropy, and radiative cooling time. We define the cool-core radius as the turnover radius in the ICM temperature profile and investigate the relation between the cool-core radius and the properties of the host galaxy clusters. In our sample, we observe that the radiative cooling time of the ICM at the cool-core radius exceeds 10\,Gyr, with RXCJ1504.1-0248 exhibiting a radiative cooling time of $32^{+5}_{-11}$\,Gyr at its cool-core radius. These results indicate that not only radiative cooling but also additional mechanisms such as gas sloshing may play an important role in determining the size of cool cores. Additionally, we might find that the best-fit relation between the cool-core radius and the cluster mass ($M_{500}$) is consistent with a linear relation. Our findings may suggest that cool cores are linked to the evolution of their host galaxy clusters.
\end{abstract}



%%%%%%%%%%%%%%%%%%%%%%%%%%%%%%%%%%%%%%%%%%%%%%%%%%%%%%%%%%%%
\section{Introduction}
\label{sec:intro}

Galaxy clusters contain a large amount of diffuse, hot X-ray emitting gas known as the intracluster medium (ICM, $T \sim 10^{7-8}$\,K), which is trapped and thermalized in the deep gravitational potential well dominated by dark matter. Because of the high temperature of the ICM, the ICM emits X-ray radiation through thermal bremsstrahlung. Since the X-ray emissivity of the ICM is proportional to the square of the electron density, the ICM loses its thermal energy by X-ray radiation faster in the central region of galaxy clusters than in the outskirts.

The radiative cooling time of the ICM at the centers of galaxy clusters exhibiting strong X-ray surface brightness peaks is much shorter than the inferred age of galaxy clusters\footnote{7.7\,Gyr ($z \sim 1$) is often adopted as the age of low-$z$ galaxy clusters.} \citep[e.g.,][for a review]{Peterson2006}. Thus, it was expected that runaway cooling occurs and triggers massive star formation in the centers of galaxy clusters \citep[][]{Fabian1994}. However, this expectation is inconsistent with observational evidence. Neither such expected massive star formation nor a large amount of cooled ICM that serves as fuel for star formation is found \citep[e.g.,][]{Tamura2001, Peterson2001, ODea2008, McDonald2011, McDonald2018}. This inconsistency indicates that runaway cooling must be suppressed, meaning the presence of heating sources.

Although a heating source is required, X-ray observations have revealed that the temperature of the ICM drops toward the cluster center. The ICM temperature at the center is measured at a few keV, corresponding to $30 - 50$\,\% of the peak value of the ICM temperature profile \citep[e.g.,][]{Vikhlinin2005, Simionescu2011}. Such a region, consisting of cooler ICM (or cooling ICM), is referred to as a cool core \citep{Molendi2001}. The presence of cool cores indicates that not only the cooling of the ICM is still dominant in cool cores but also the heating is balanced with the cooling of the ICM at least within a specific region. Therefore, exploring cool-core systems in galaxy clusters is important to understand not only the origin of cool cores but also the evolution of cool cores under conditions of the balance between cooling and heating.

Cool cores are characterized by a significant drop in the ICM temperature toward the cluster center \citep{Molendi2001}. However, various alternative observables of the ICM have been used to identify cool cores: central electron density \citep{Hudson2010, Barnes2018}, central radiative cooling time \citep[e.g.,][]{Bauer2005, Hudson2010, Wang2023}, central entropy profile \citep[e.g.,][]{Bauer2005, Hudson2010}, classical mass deposition rate \citep[e.g.,][]{Chen2007}, and X-ray luminosity ratio \citep[e.g.,][]{Sayers2013, Shitanishi2018}. These observables have also been used to distinguish between cool-core and non-cool-core clusters. For instance, \cite{Su2020} used the central radiative cooling time of the ICM as an indicator to identify cool-core clusters from a sample of galaxy clusters obtained from cosmological simulations such as IllustrisTNG. Additionally, \cite{Barnes2018} investigated the cool-core faction in a sample of galaxy clusters from IllustrisTNG using six parameters: central electron density, radiative cooling time, entropy profile, X-ray concentration parameters within a certain or fiducial radius, and cuspiness parameter of X-ray morphology. \cite{Lagana2019} also investigated the optimal parameters for identifying cool cores using some observables including the cuspiness of the gas density profile, central gas density, and properties of the brightest cluster galaxies (BCG).

\cite{Hudson2010} presented an in-depth study of the ICM properties in the central regions of galaxy clusters listed in a catalog of the extended HIghest X-ray FLUx Galaxy Cluster Sample \citep[HIFLUGCS:][]{Reiprich2002}. They applied 16 cool-core diagnostics parameters to their sample to find the most appropriate parameter for characterizing cool-core clusters. They found that the radiative cooling time at the center and cuspiness serve as the most effective indicators for low-$z$ and high-$z$ cool-core clusters, respectively. In addition, they introduced two categories of cool cores: strong and weak cool cores, based on the radiative cooling time at the center. Strong cool cores are defined as those with a radiative cooling time at the center shorter than 1\,Gyr, while weak cool cores are identified when a radiative cooling time ($t_{\rm cool}$) at the center falls within the range of 1\,Gyr $ < t_{\rm cool} < $7.7\,Gyr.

A universal form in the ICM temperature profile has been reported \citep[e.g.,][]{Allen2001, Vikhlinin2005, Sanderson2006}. \cite{Allen2001} analyzed the ICM temperature profiles scaled by $r_{2500}$\footnote{We adopt $M_{\Delta}$ as the mass enclosed within a sphere of radius $r_{\Delta}$ whose mean density is $\Delta$ times the critical density of the Universe at the redshift of the galaxy cluster.} for six cool-core clusters and found an approximately universal form in the scaled temperature profiles. 
\cite{Vikhlinin2005} measured the ICM temperature profiles from the central region to the outskirts in 13 nearby, relaxed galaxy clusters and groups, and revealed a similar trend in the temperature profiles scaled by $r_{180}$. \cite{Sanderson2006} also measured the ICM temperature profiles in 20 galaxy clusters and found a possible universal form in the temperature profiles scaled by $r_{500}$. 

Such previous studies aim to investigate possible relations between the temperature profile and the total mass of galaxy clusters. However, both cooling and heating processes have a significant impact on cool cores, indicating that baryon physics in the centers of galaxy clusters likely plays an important role in shaping the observed characteristics of cool cores. Therefore, it is essential to focus on cool-core systems and characterize cool cores by the ICM temperature.

In this paper, we aim to characterize cool cores in our sample using the cool-core radius that is defined as the turnover radius in the ICM temperature profile. In addition, we aim to study possible relations between the cool-core radius and the properties of the host galaxy clusters, including the ICM thermodynamic properties and cluster mass. To this end, we first analyze the X-ray spectra of the ICM extracted from annular regions determined by the morphology of X-ray surface brightness and X-ray photon counts. Then, we measure the radial profiles of the ICM thermodynamic properties and determine the cool-core radius by analyzing the ICM temperature profile. Finally, we measure thermodynamic perturbations in the ICM to constrain the turbulent velocity of the ICM within the cool cores in our sample.

This paper is organized as follows. Section~\ref{sec:sample} provides a brief summary of our sample and their properties. Section~\ref{sec:obs} presents detailed information regarding \Chandra ~observations and data reduction procedures. Section~\ref{sec:ana} describes the X-ray spectral analysis and measurements of the radial profiles of the ICM thermodynamic properties. Section~\ref{sec:discussion} presents the study of possible relations between the cool-core radius and the properties of the host galaxy clusters and the analysis of thermodynamic perturbations in the ICM. Finally, a summary is given in Section~\ref{sec:summary}

Throughout the paper, we assume $\Omega_{\rm m}=0.3$, $\Omega_{\rm \Lambda}=0.7$, and the Hubble constant of $H_{0} = 70\,$km\,s$^{-1}$\,Mpc$^{-1}$. Unless stated otherwise, quoted errors correspond to 1\,$\sigma$ uncertainties.

\section{Sample}
\label{sec:sample}

In this paper, we focus on four galaxy clusters: RXCJ1504.1-0248, Abell\,3112, Abell\,4059, and Abell\,478. These galaxy clusters are included in the HIFLUGCS catalog \citep{Reiprich2002}. According to \cite{Hudson2010}, they are classified as strong cool-core clusters based on the radiative cooling time of the ICM at the center. From the sample of the strong cool-core clusters, these four clusters are selected based on (1) relatively hot gas \citep[virial temperature of $kT_{\rm vir} > 4$\,keV;][]{Hudson2010}, (2) cooling, low-entropy gas at the center \citep[cooling time of $t_{\rm cool} < 0.6$\,Gyr or entropy of $K_{0} < 10$\,keV\,cm$^2$;][]{Hudson2010}, (3) high-contrast sloshing \citep[all perturbations with entropy contrast of $K_{\rm neg}/ K_{\rm pos} > 1.3$;][]{Ueda2021}. Thus, they are characterized as relatively massive clusters hosting sloshing cool cores where a strong heating source(s) is required. Here, we provide a brief summary of previous studies on each galaxy cluster. The values of $M_{500}$ and $r_{500}$ mentioned below are extracted from the Meta-Catalogue of X-ray detected Clusters of galaxies \citep[MCXC;][]{Piffaretti2011}.

\subsection{RXCJ1504.1-0248}

This cluster is located at a redshift of $z = 0.2153$ and is known as one of the most massive galaxy clusters with $M_{500} = 12.47 \times 10^{14}$\,\MO ~($r_{500} = 1.52 $\,Mpc). This cluster hosts an extreme cool core characterized by a short radiative cooling time \citep[e.g.,][]{Bohringer2005, HlavacekLarrondo2011}. \cite{Bohringer2005} analyzed the X-ray surface brightness profile using a $\beta$-model \citep{Cavaliere1976, Cavaliere1978, Ettori2000} and found that the core radius of the $\beta$-model is $\sim 30$\,$h_{70}^{-1}$\,kpc\footnote{$h_{70}$ denotes $h_{70} = H_{0} / (70 $\,km\,s$^{-1}$\,Mpc$^{-1}$) where $H_{0}$ is a Hubble constant used in the literature.}, which is significantly smaller than the cooling radius of $\sim 140$\,kpc which is defined as the position at a radiative cooling time of $10$\,Gyr. Furthermore, \cite{Bohringer2005} found a significant drop in the ICM temperature profile toward the cluster center. The ICM temperature at the center is measured to be below 5\,keV, while the temperature of the ambient ICM is $\sim 10.5$\,keV. \cite{HlavacekLarrondo2011} reported that there is no obvious X-ray point source associated with the BCG. \cite{Giacintucci2011} found the presence of a radio mini-halo in the central region of this cluster.

\subsection{Abell\,3112}

This cluster is located at a redshift of $z = 0.075$ and has a powerful radio source, PKS 0316-444, in the center. \cite{Takizawa2003} measured the radial profiles of the ICM thermodynamic properties, including temperature, abundance, electron density, pressure, and radiative cooling time with \Chandra, and found a temperature drop toward the cluster center. \cite{Bulbul2012} also studied the temperature profile of the ICM from the center to the outskirts. The ICM temperature at the center is measured at $\sim 3.4$\, keV, while the peak value in the temperature profile is measured at $\sim 5.1$\,keV, which is consistent with that reported by \cite{Ezer2017}. The mass of this cluster is estimated at $M_{\rm 500} = 4.39 \times 10^{14}$\,\MO ~($r_{500} = 1.13$\,Mpc), which is consistent with that estimated by \cite{Nulsen2010} and  \cite{Bulbul2012}.

\subsection{Abell\,4059}
\label{sec:A4059}

This cluster is located at a redshift of $z = 0.049$. The ICM temperature profile shows a temperature drop from $\sim 4$\,keV at $\sim 100$\,kpc away from the center to $\sim 2$\,keV at the center \citep[e.g.,][]{Huang1998, Choi2004, Reynolds2008, Mernier2015}. \cite{Lagana2019} conducted a detailed study of spatial distributions of the ICM temperature, entropy, pressure, and abundance, respectively. The mass of this cluster is estimated as $M_{\rm 500} = 2.67 \times 10^{14}$\,\MO ~($r_{500} = 0.96$\,Mpc), which is consistent with that estimated by \cite{Vikhlinin2009}. \cite{Reynolds2008} reported a slight dip in the pressure profile at $r \sim 15$\,kpc, which is likely associated with an X-ray cavity caused by AGN feedback. Thus, it is considered that a lack of thermal pressure may be supported by its non-thermal pressure.

\subsection{Abell\,478}

This cluster is located at a redshift of $z = 0.088$. \cite{Sun2003} analyzed the central 500\,kpc region with \Chandra ~and reported the radial profiles of the ICM thermodynamic properties, finding that the peak value in the temperature profile is $\sim 8.5$\,keV, while the temperature at the center is $\sim 3$\,keV. These results are in agreement with those measured by \cite{Pointecouteau2004} and \cite{Vikhlinin2005}. X-ray cavities are observed in the central 15\,kpc region, with two weak and small ($\sim$ 4\,kpc) radio lobes spatially associated with the X-ray cavities \citep{Sun2003}. The mass of this cluster is estimated as $M_{\rm 500} = 6.42 \times 10^{14}$\,\MO  ~($r_{500} = 1.28$\,Mpc).

\section{Observation and data reduction}
\label{sec:obs}

We analyzed archival X-ray data of the sample taken with the Advanced CCD Imaging Spectrometer \citep[ACIS;][]{Garmire2003} on board the \Chandra ~X-ray Observatory. This paper employs a list of Chandra datasets, obtained by the Chandra X-ray Observatory, contained in~\dataset[DOI:10.25574/cdc.223]{https://doi.org/10.25574/cdc.223}. The observation identification numbers (ObsIDs) and corresponding information of \Chandra ~observations in this study are summarized in Table~\ref{tab:list}. We used the versions of 4.11 and 4.8.3 for \Chandra ~Interactive Analysis of Observations \cite[CIAO;][]{Fruscione2006} and the calibration database (CALDB), respectively. To ensure data quality, we examined the light curve of each dataset using the {\tt lc\_clean} task in CIAO, filtering flare data. The blanksky data provided by the CALDB were adopted as background data for the spectral analysis. Point sources were identified by the {\tt wavdetect} task in CIAO and were subsequently masked. We extracted X-ray spectra of the ICM from each dataset using the {\tt specexctract} task in CIAO and combined them after making individual spectrum, response, and ancillary response files for the spectral fitting. We used {\tt XSPEC} version 12.11.0f \citep{Arnaud1996} and the atomic database ({\tt AtomDB}) for plasma emission modeling version 3.0.9 in the X-ray spectral analysis, assuming that the ICM is in collisional ionization equilibrium \citep{Smith2001, Foster2012}. The abundance table of \cite{Anders1989} was used in {\tt XSPEC}. Here, the abundance of a given element is defined as $Z_{i} = (n_{i, {\rm obs}}/n_{\rm H, obs}) / (n_{i, \odot} / n_{\rm H, \odot})$, where $n_{i}$ and $n_{\rm H}$ represent the number densities of the $i$th element and hydrogen, respectively. The iron abundance of the ICM is used to represent the ICM metal abundance, such that the abundance of other elements is tied to the iron abundance as $Z_{i} = Z_{\rm Fe}$ \citep{Ueda2021}.

%%%%%%%%%%%%%%%%%%%%%%%%%%%%
\begin{table*}[htbp]
\centering
\caption{
Summary of our sample: cluster name, redshift, net exposure time, physical scale, and datasets taken with \Chandra.
}\label{tab:list}
\begin{tabular}{lcccl}
\hline\hline	
Cluster				& Redshift		& Expo. time (ksec)	& Scale (kpc arcsec$^{-1}$)      & ObsID							\\ \hline
RXCJ1504.1-0248 		& 0.215	    	& 161.7			& 3.51                          		& 4935, 5793, 17197, 17669, 17670		\\
A3112          			& 0.075	   	& 133.7			&1.43                          		& 2216, 2516, 6972, 7323, 7324, 13135	\\
A4059           			& 0.049	   	& 120.0			& 0.96                          		& 897, 5785						\\
A478            			& 0.088	   	& 50.1			& 1.65                          		& 1669, 6102						\\
\hline
\end{tabular}
\end{table*}
%%%%%%%%%%%%%%%%%%%%%%%%%%%%

%%%%%%%%%%%%%%%%%%%%%%%%%%%%
\begin{table}[htbp]
    \begin{center}
    \caption{
    Sky coordinates (J2000.0) of the center, the position angle (PA), and the axis ratio (AR) of the ellipse model for the X-ray surface brightness distribution of the sample extracted from \cite{Ueda2021}.
    }\label{tab:sky}
    \begin{tabular}{lcccccc}
    \hline\hline	
    Cluster				& R.A.		& Decl.	& PA\tablenotemark{a}	    & AR	\\ 
    					&			&		& (deg)					& \\ \hline
    RXCJ1504.1-0248	& 15:04:07.48	& -02:48:17.25  & 156 & 0.78\\
    A3112			& 03:17:57.67	& -44:14:17.44  & 100 & 0.75\\
    A4059			& 23:57:00.79	& -34:45:33.44  &  70 & 0.88\\
    A478			& 04:13:25.15	& 10:27:54.94   & 132 & 0.73\\
    \hline
    \end{tabular}
    \end{center}
    \tablenotetext{a}{
     Position angle measured south of east.}
    \end{table}
%%%%%%%%%%%%%%%%%%%%%%%%%%%%

\section{Analysis and Results}
\label{sec:ana}

To determine the cool-core radius for the sample, our initial goal is to obtain and analyze the ICM temperature profile of each cluster. We define annular regions for X-ray spectral analysis based on the morphology of the X-ray surface brightness of the sample. Next, we extract X-ray spectra of the ICM from each defined annular region and carry out spectral analysis to measure the ICM thermodynamic properties. Then, we perform fitting of the ICM temperature profile to determine the cool-core radius. We also study the radial profiles of the ICM electron number density, pressure, entropy, and radiative cooling time.

\subsection{X-ray imaging analysis}
\label{sec:reg}

\begin{figure*}
    \begin{center}
     \includegraphics[width=7.5cm]{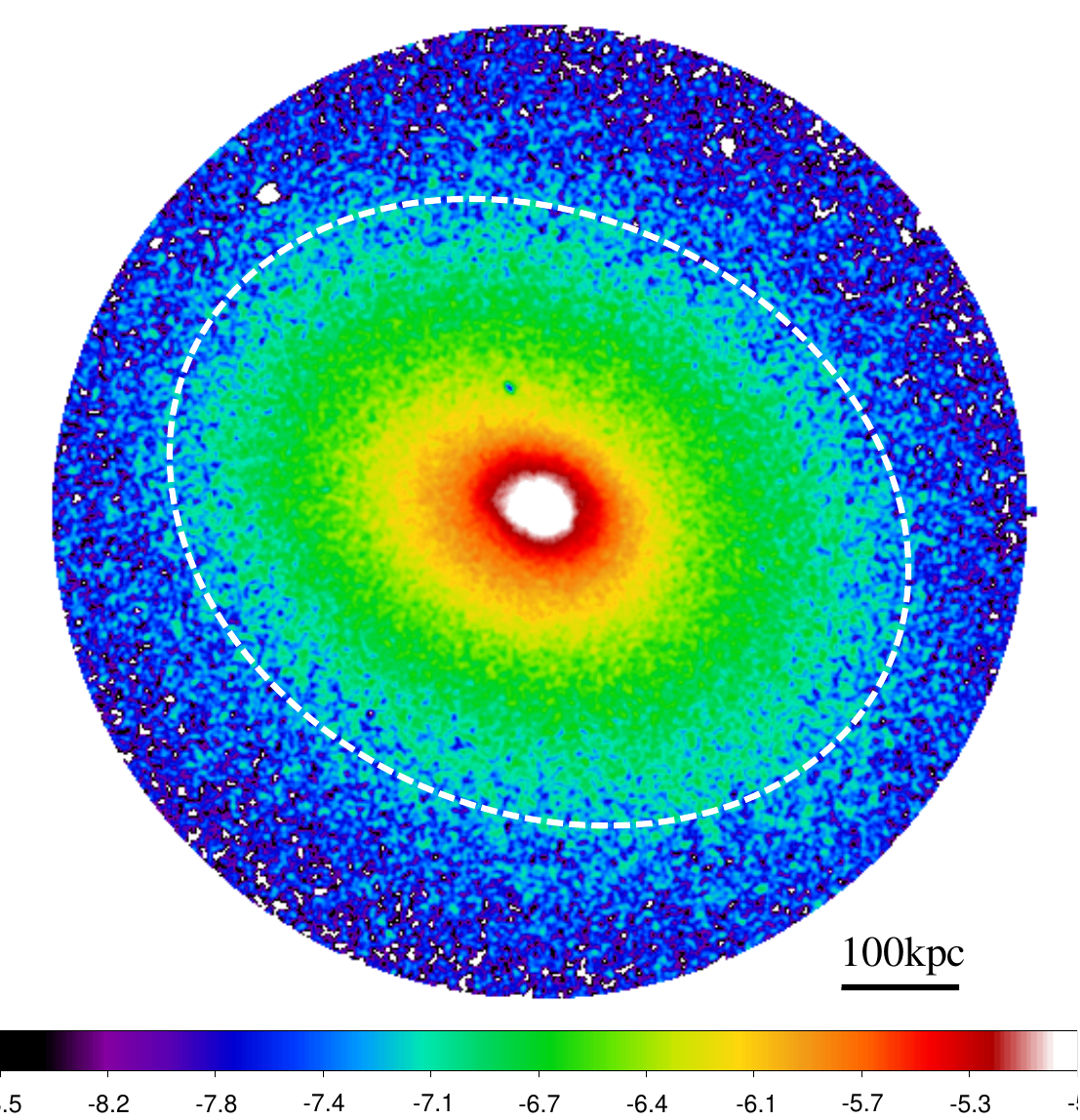}
     \includegraphics[width=7.5cm]{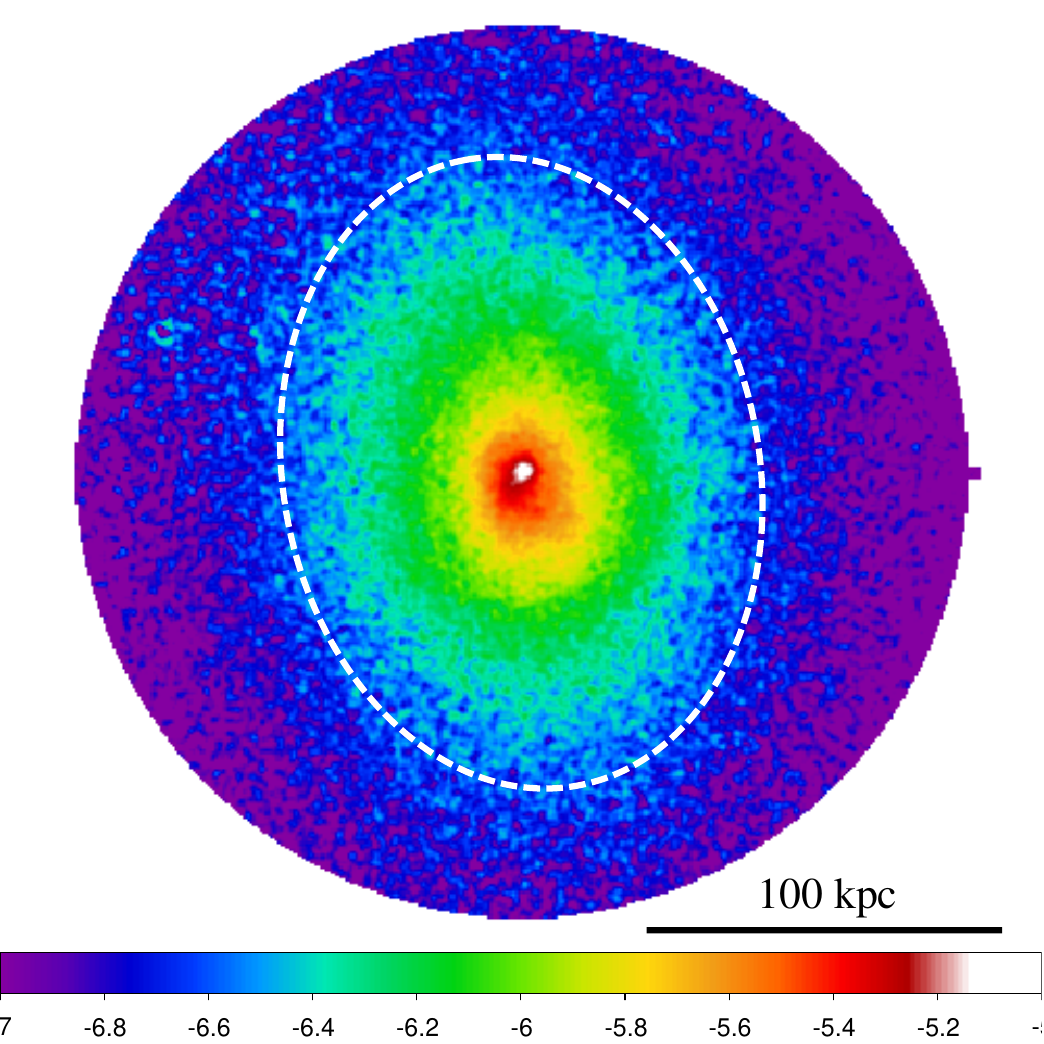}
     \includegraphics[width=7.5cm]{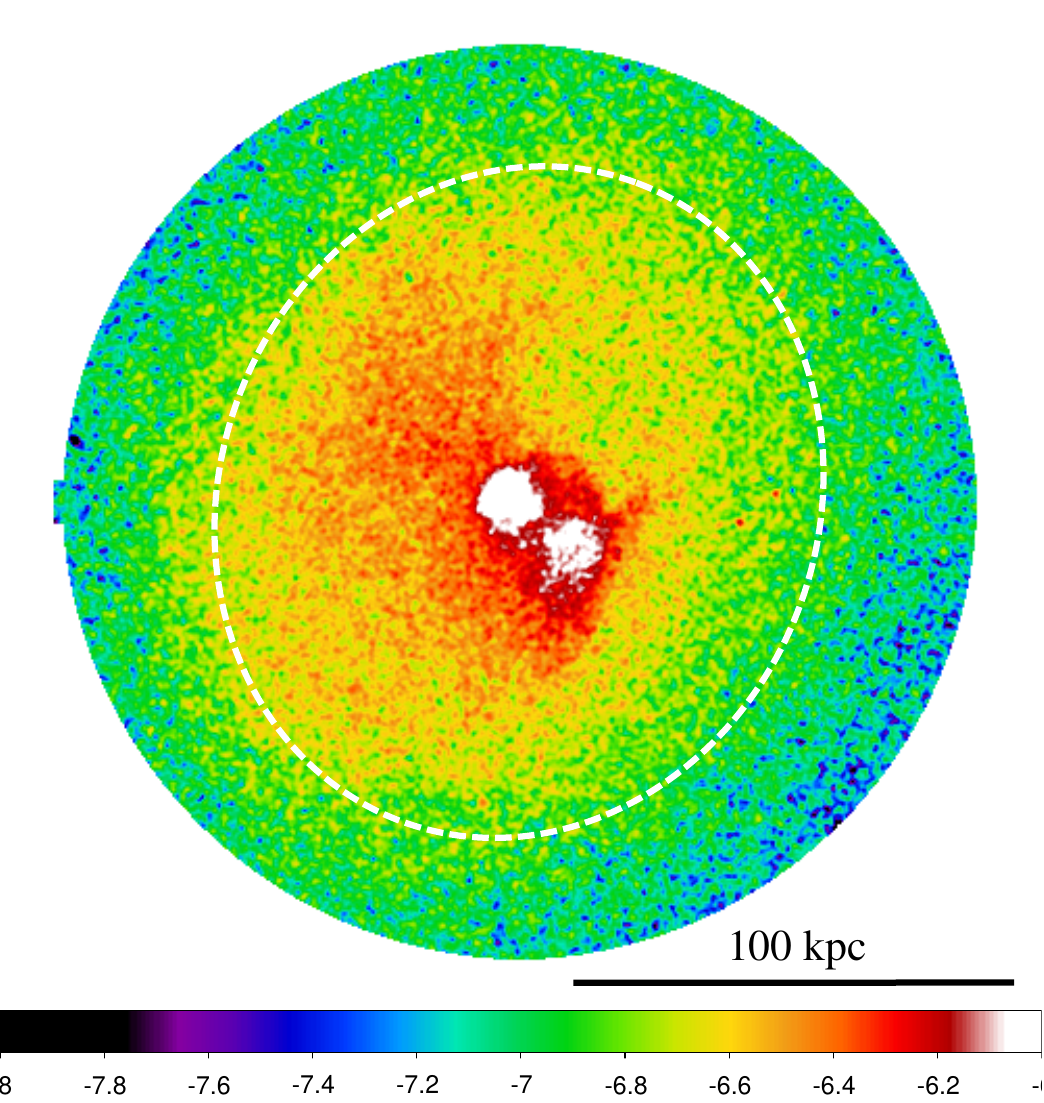}
     \includegraphics[width=7.5cm]{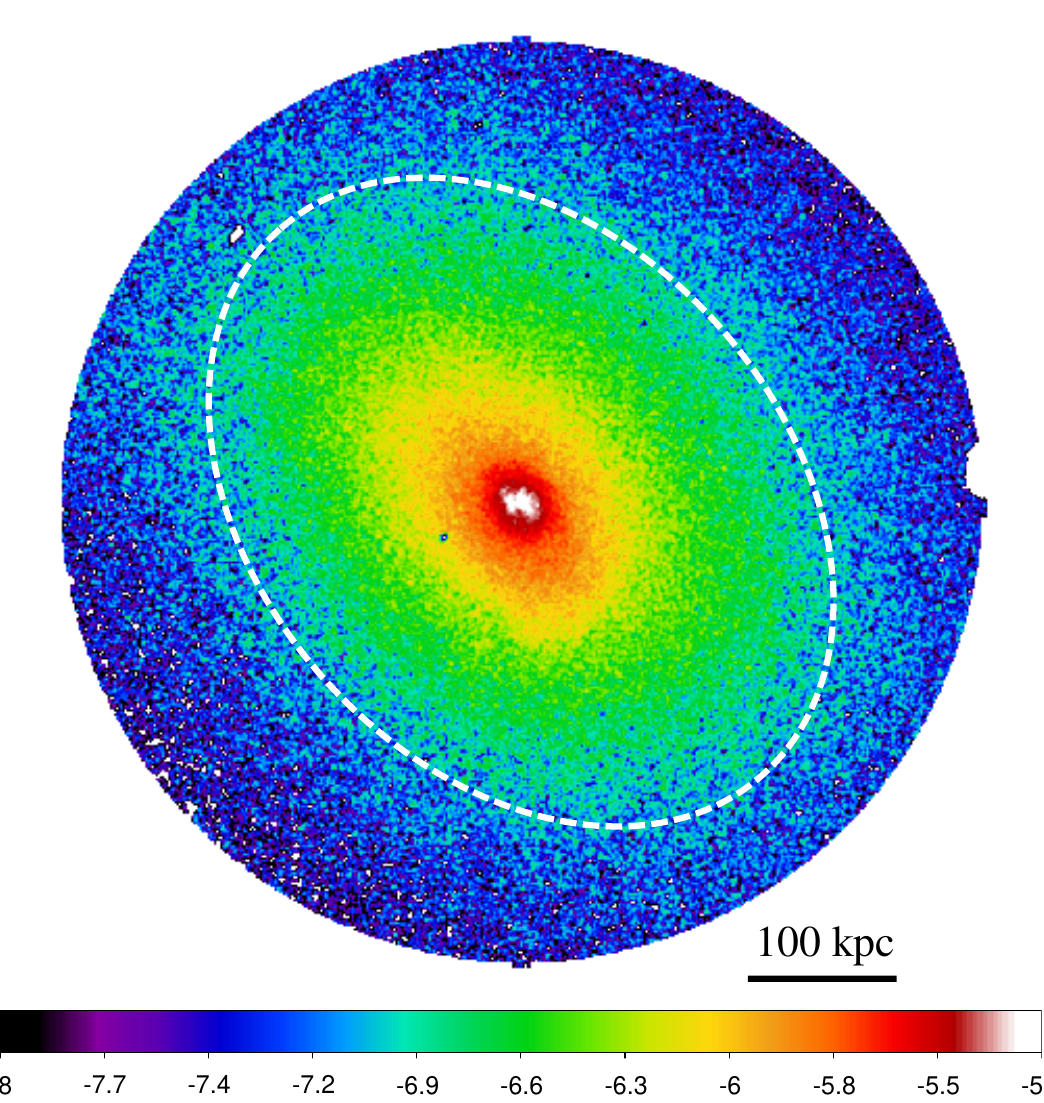}
    \end{center}
\caption{X-ray surface brightness of the sample: RXCJ1504.1-0248 (top left), A3112 (top right), A4059 (bottom left), and A478 (bottom right). The X-ray surface brightness in the $0.5 - 7.0$\,keV band is shown on a logarithm scale in units of photon\,sec$^{-1}$\,arcsec$^{-2}$\,cm$^{-2}$. This image is smoothed with a Gaussian kernel with $2.3''$ FWHM. A dashed, white ellipse shows the cool-core radius, $r_{\rm cool}$, derived from the best-fit results of the analysis of the ICM temperature profile.
}
\label{fig:image}
\end{figure*}

Figure~\ref{fig:image} shows the X-ray surface brightness images of the sample in the $0.5 - 7.0$\,keV band taken with \Chandra ~after subtracting the background and correcting the exposure time. Since the morphology of their X-ray surface brightness appears axial symmetric \citep[e.g.,][]{Ueda2021}. we adopt the same elliptical model as that used in \cite{Ueda2021} to define annular regions for spectral analysis. The position of the center, the position angle, and the axis ratio of the elliptical model for each cluster are summarized in Table~\ref{tab:sky}. Note that \cite{Ueda2021} calculated a mean surface brightness by applying the concentric ellipse fitting algorithm constructed by \cite{Ueda2017}, by minimizing the variance of the X-ray surface brightness relative to the ellipse model. To determine the width of each radial bin, we ensure that the net photon counts for each bin falls within the range of $5000 - 10000$ in the $0.4 - 7.0$\,keV band for better statistics. For several smaller regions near the center and outer regions, we adjust the net photon counts in the range of $1500 - 5000$. The detailed information regarding our region selection is summarized in Appendix~\ref{sec:apx_reg} (see Table~\ref{tab:annulus_regions}).

%%%%%%%%%%%%%%%%%%%%%%%%%%%%
%%%%%%%%%%%%%%%%%%%%%%%%%%%%

\subsection{X-ray spectral analysis}
\label{sec:spec}

We extract X-ray spectra of the ICM from each elliptical annulus defined in Section~\ref{sec:reg}. The X-ray spectra in the $0.4 - 7.0$\,keV band are analyzed using the model of {\tt phabs * apec} in {\tt XSPEC}. The redshift of each cluster is fixed to the value in Table~\ref{tab:list}, which is taken from NASA/IPAC Extragalactic Database (NED)\footnote{http://ned.ipac.caltech.edu/}. The column densities of the Galactic absorption (i.e., $N_{\rm H}$) toward RXCJ1504.1-0248 and A3112 are fixed to the values measured by \cite{HI4PI2016}, respectively. However, for A4059 and A478, \cite{Choi2004} and \cite{Sun2003} reported that $N_{\rm H}$ varies with radius and is systematically larger than that derived from \cite{HI4PI2016}. In fact, the measured values of $N_{\rm H}$ for A4059 and A478 in our spectral analysis are a factor of 2 larger than those of \cite{HI4PI2016}, respectively, and consistent with the previous measurements \citep[][]{Choi2004, Sun2003}, respectively. Therefore, we allow $N_{\rm H}$ for A4059 and A478 to vary in the spectral analysis.

\begin{figure*}
    \begin{center}
        \includegraphics[width=8.5cm]{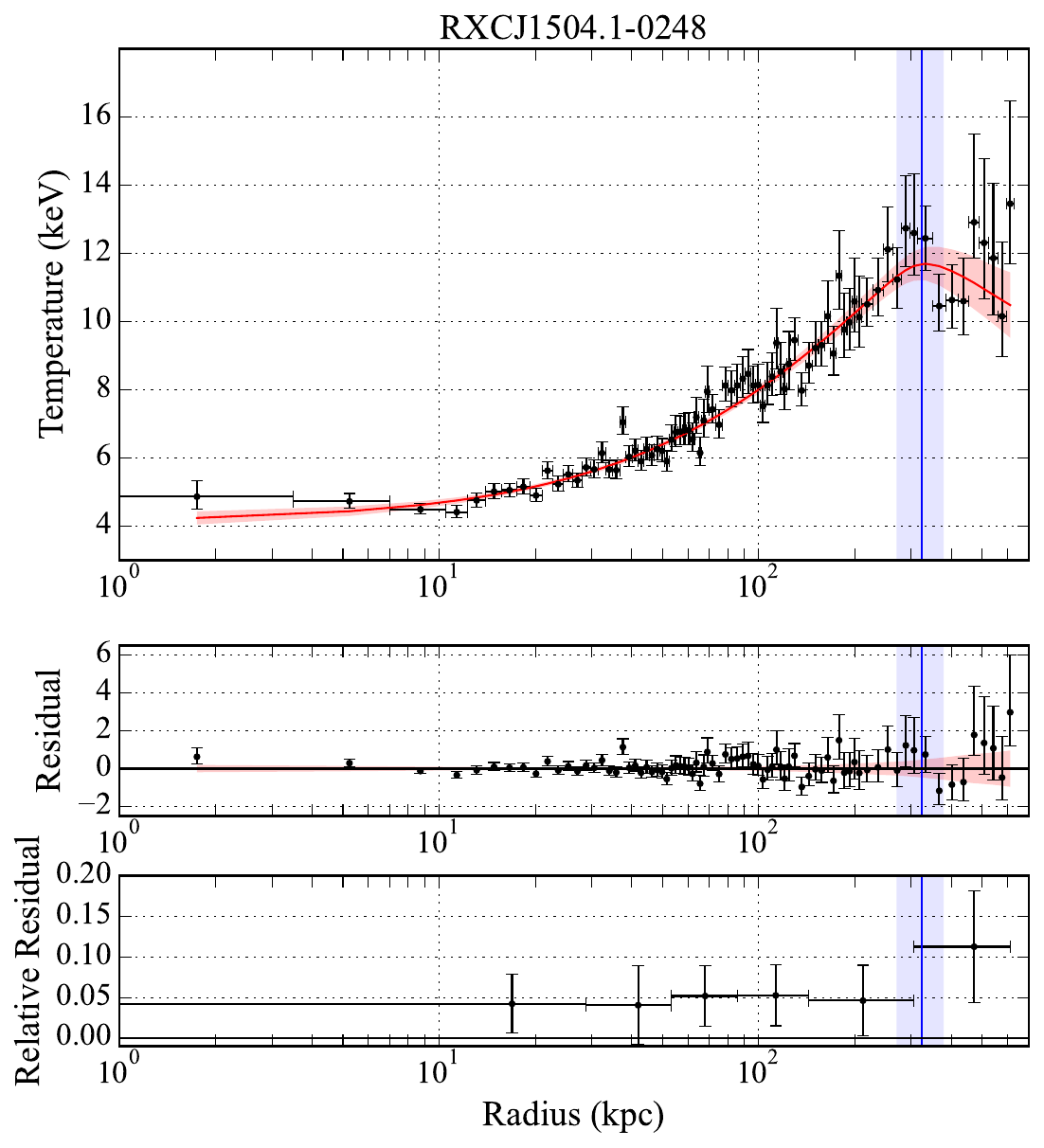}
        \includegraphics[width=8.5cm]{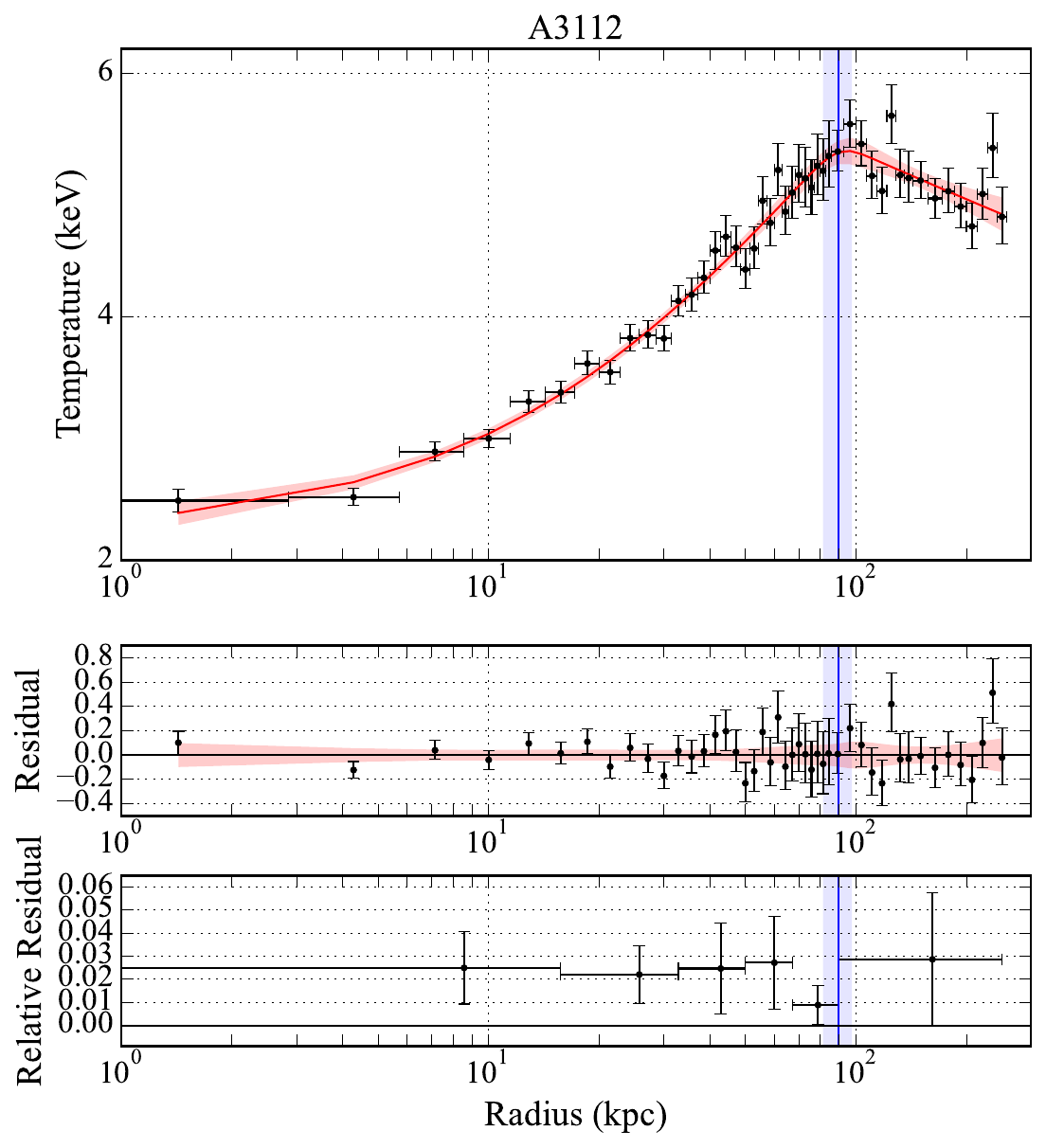}
        \includegraphics[width=8.5cm]{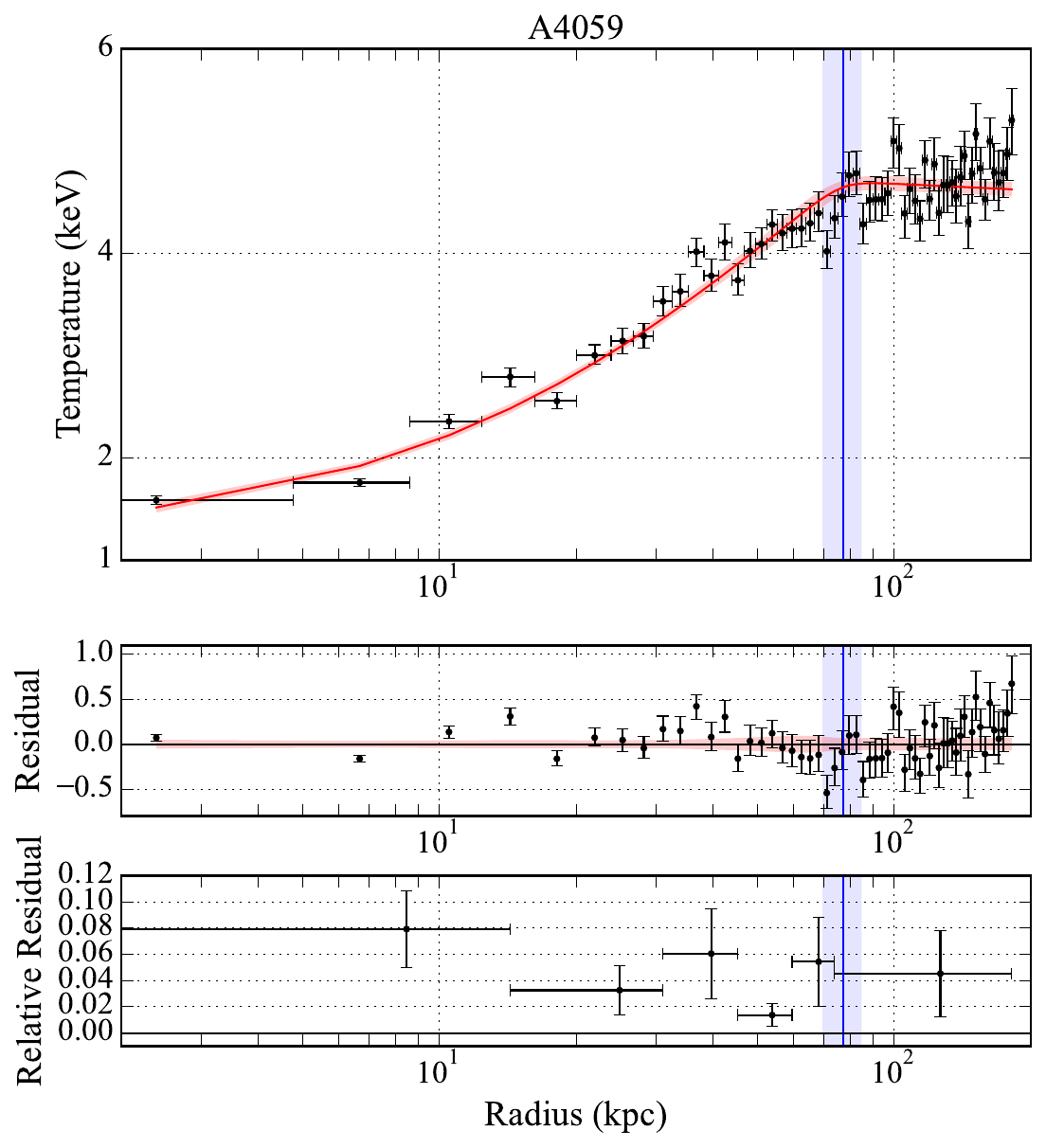}
        \includegraphics[width=8.5cm]{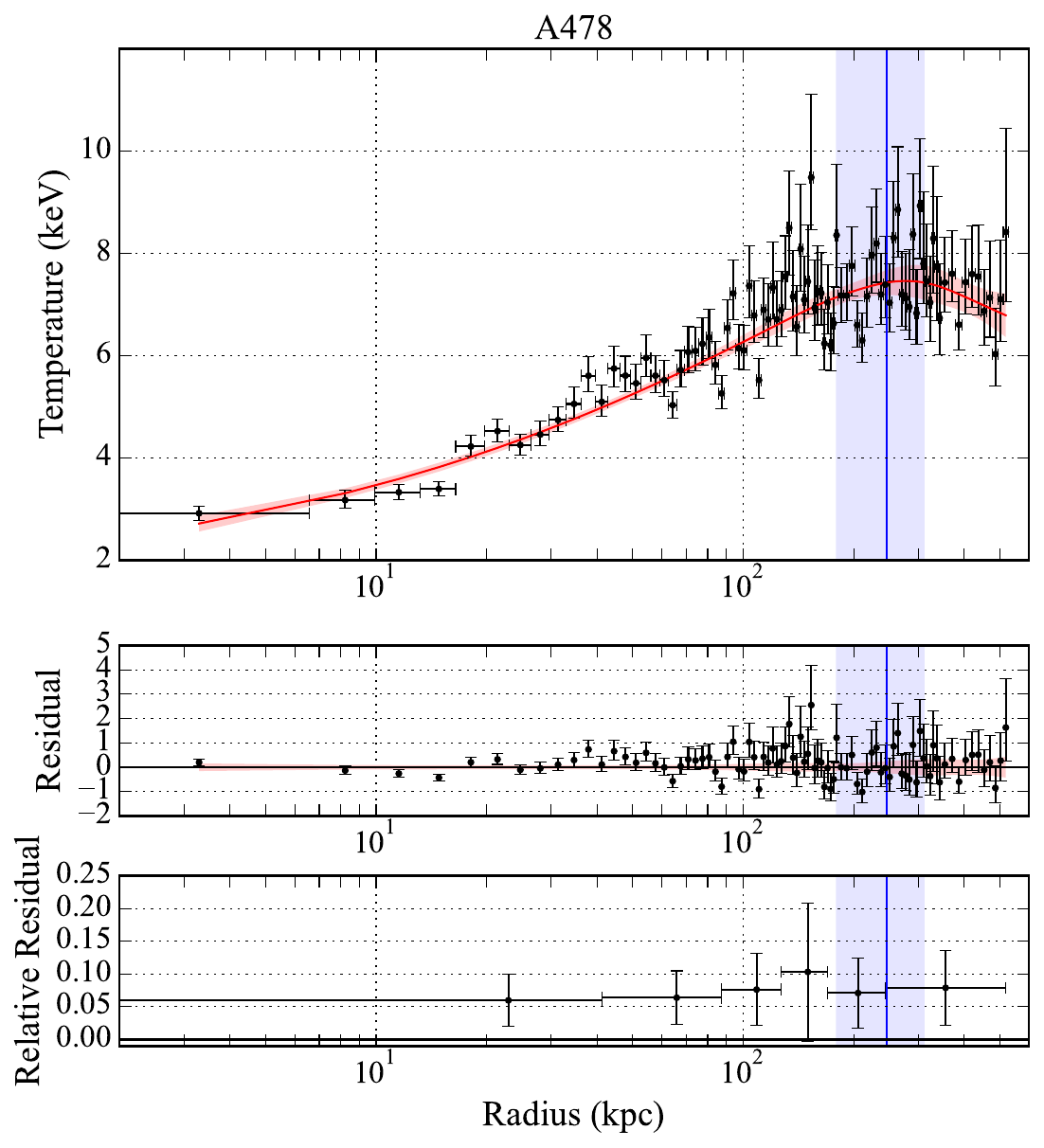}
    \end{center}
\caption{ICM temperature profiles of the sample: RXCJ1504.1-0248 (top left), A3112 (top right), A4059 (bottom left), and A478 (bottom right). In each panel, the black error bars show the observed values. The red dashed line and shaded region display the best-fit profile and $1\sigma$ uncertainty, respectively. The blue solid line and shaded region correspond to the best-fit cool-core radius $r_{\rm cool}$ and $1\sigma$ uncertainty, respectively. In the middle row of each panel, the residuals between the observed and the best-fit profiles are shown. The mean absolute relative residuals are also displayed in the bottom row of each panel.
}
\label{fig:kT_fitting}
\end{figure*}

The observed radial profiles of the temperature and electron number density for the sample are shown in Figures~\ref{fig:kT_fitting} and \ref{fig:nD_fitting}, respectively. To estimate the ICM electron number density, we assume a line-of-sight length of $L$/1\,Mpc. Additionally, the observed radial profiles of the ICM metal abundance are presented in Appendix (Figure~\ref{fig:abund_plot}). We also calculate the ICM pressure $p_{\rm e}$ and entropy $K_{\rm e}$ as
\begin{equation}
    p_{\rm e} = kT \times n_{\rm e}
    \label{eqt:P}
\end{equation}
and 
\begin{equation}
    K_{\rm e} = kT \times n_{\rm e}^{-\frac{2}{3}},
    \label{eqt:K}
\end{equation}
respectively, where $kT$ is the ICM temperature and $n_{\rm e}$ is the ICM electron number density. The radial profiles of the pressure and entropy are shown in Figures~\ref{fig:P_fitting} and \ref{fig:K_fitting}, respectively. In addition, following the approach of \cite{McDonald2019}, we calculate the radiative cooling time of the ICM, $t_{\rm cool}$, as
\begin{equation}
    t_{\mathrm{cool}} = \frac{3}{2} \frac{(n_{\mathrm e} + n_{\mathrm p}) kT}{n_{\mathrm e} n_{\mathrm p} \Lambda(T, Z)},
    \label{eqt:tcool}
\end{equation}
where $n_{\mathrm p}$ is the proton number density, and $\Lambda(T, Z)$ is the cooling function. To convert from $n_{\mathrm e} n_{\mathrm p}$ to $n_{\mathrm e}$, \cite{McDonald2019} assumed $n_{\mathrm e}/n_{\mathrm p} = 1.196$, based on the discussion by \cite{Markevitch2007b}, namely, $n_{\mathrm e}/n_{\mathrm p} = 1 + 2x + x_{\mathrm eh}$, where $x \equiv n_{\rm He}/n_{\mathrm p}$ represents the helium abundance, and $x_{\mathrm eh}$ represents electrons from elements heavier than helium. The contribution of $x_{\mathrm eh}$ can be negligible, as it is $x_{\mathrm eh} \approx 0.005$ assuming an ICM abundance of $0.3 - 0.5$\,solar compared to $x = 0.098$ from \cite{Anders1989} with 1\,solar abundance. Therefore, we adopt the same approach as \cite{McDonald2019} in calculation of the radiative cooling time. We compute the cooling function using the {\tt pyatomdb} task in {\tt AtomDB} \citep{Smith2001, Foster2012} with the best-fit parameters of the ICM temperature and abundance in each radial bin. The radial profiles of the radiative cooling time for the sample are presented in Figure~\ref{fig:tcool_plot}. To show all the radial profiles, we use the distance from the center along the direction of the major axis of the elliptical model as the values on the horizontal axis.\footnote{In this paper, we adopt $r = \frac{r_\mathrm{MAJ,\, inner} + r_\mathrm{MAJ,\, outer}}{2}$ as the distance from the center, where $r_\mathrm{MAJ,\, inner} $ and $ r_\mathrm{MAJ,\, outer}$ denote the major axes of the inner and outer annuli, respectively.} Note that we here show the radial profiles of the ICM properties derived from projected spectral analysis, not spectral deprojection analysis. 

%%%%%%%%%%%%%%%%%%%%%%%%%%%%
%%%%%%%%%%%%%%%%%%%%%%%%%%%%

\subsection{ICM temperature profiles}
\label{sec:temp_fitting}

Figure~\ref{fig:kT_fitting} shows the observed temperature profile of each galaxy cluster. In good agreement with previous studies, such as \cite{Bohringer2005} for \rxcj, \cite{Takizawa2003} for A3112, \cite{Reynolds2008} for A4059, and \cite{Sanderson2005} for A478, all galaxy clusters in the sample exhibit a significant drop in the observed ICM temperature profiles toward the cluster center. 

Several empirical models have been proposed to represent the ICM temperature profile from the center to the outskirts \citep[e.g.,][]{Allen2001, Voigt2002, Kaastra2004, Vikhlinin2006, OSullivan2017}. For instance, \cite{Kaastra2004} modeled the ICM temperature profile taking into account the temperature at the center, the peak temperature, and the cooling radius that represents a characteristic radius in the temperature profile. However, there is still room for improvement in their model to account for temperature decrease toward the outskirts.

Motivated by \cite{Kaastra2004} and \cite{OSullivan2017}, we extend their models to account for the temperature profile from the center to the outer region. Therefore, our model can be described as
\begin{equation}
    \label{eqt:kT}
    T(r) = T_{\rm center} + 2 \times(T_{\rm peak} - T_{\rm center})(\frac{x(r)}{1+x(r)})
\end{equation}
\begin{equation}
    x(r) =
    \begin{cases}
    (r / r_{\rm cool})^{ \alpha_1} & : r < r_{\rm cool} \\
    (r / r_{\rm cool})^{ \alpha_2} & : r \geq r_{\rm cool},
    \end{cases}
\end{equation}
where $T_{\rm center}$ is the ICM temperature at the center, $T_{\rm peak}$ is the peak temperature, $r_{\rm cool}$ is the turnover radius in the temperature profile (i.e., the peak position; $T(r_{\rm cool}) = T_{\rm peak}$), and $\alpha_1$ and $\alpha_2$ represent the slopes of the temperature profile within $r_{\rm cool}$, and beyond $r_{\rm cool}$, respectively. In this paper, we define the cool-core radius as $r_{\rm cool}$. 

We fit the observed temperature profile shown in Figure~\ref{fig:kT_fitting} with our model using affine-invariant Markov Chain Monte Carlo (MCMC) sampling \citep{Goodman2010} implemented by the {\tt emcee} python package \citep{ForemanMackey2013}. The log-likelihood function for the fitting is written as
\begin{equation}
-2 \ln \mathcal{L} = \sum_{i} \frac{[y_{i} - T(r_{i})]^{2}}{\sigma_{y_{i}}^2},
\end{equation}
where $i$ runs over all radial bins in the radial profile, $y_{i}$ and $\sigma_{y_i}$ are the best-fit value and its uncertainty of the temperature profile in each radial bin, respectively, and $T(r_i)$ is the model prediction in each radial bin. We use uninformative uniform priors on $T_{\rm center}$, $T_{\rm peak}$, $r_{\rm cool}$, $\alpha_1$, and $\alpha_2$ with the following ranges: $T_{\rm center} \in (0, 30)$\,keV, $T_{\rm peak} \in (0, 30)$\,keV, $r_{\rm cool} \in (0, 500)$\,kpc, $\alpha_1 \in (0, 2)$, and $\alpha_2 \in (-2, 0)$. We sample the posterior probability distributions of the parameters ($T_{\rm center}$, $T_{\rm peak}$, $r_{\rm cool}$, $\alpha_1$, and $\alpha_2$) over the full parameter space allowed by the priors. The mean value and standard deviation of the marginalized distributions are represented as the best-fit value and its uncertainty, respectively. For each cluster, the best-fit parameters of the temperature profile are summarized in Table~\ref{tab:best}. The best-fit model of the temperature profile of each cluster is also shown in Figure~\ref{fig:kT_fitting}. 

%%%%%%%%%%%%%%%%%%%%%%%%%%%%
\begin{table*}[htbp]
    \begin{center}
    \caption{
        Parameter constraints of the temperature profile model derived from the MCMC analysis.
    }\label{tab:best}
    \begin{tabular}{lccccc}
    \hline\hline	
    Cluster				& $T_{\rm center}$ (keV)  	& $T_{\rm peak}$ (keV) 	& $r_{\rm cool}$ (kpc) 	& $\alpha_1$      	& $\alpha_2$   		\\	\hline
    RXCJ1504.1-0248	        & $4.13 \pm 0.24 $ 		&$12.00 \pm 0.56 $ 		&$325 \pm 55 $ 		&$0.97 \pm 0.11 $ 	&$-0.66 \pm 0.48 $	\\
    A3112	        			& $2.21 \pm 0.16 $ 		&$5.41 \pm 0.11 $ 		&$89.4 \pm 8.0 $ 			&$0.88 \pm 0.10 $ 	&$-0.35 \pm 0.14 $	\\
    A4059	        			& $1.21 \pm 0.13 $ 		&$4.70 \pm 0.08 $ 		&$77.1 \pm 8.7 $ 			&$0.90 \pm 0.10 $ 	&$-0.05 \pm 0.05 $	\\
    A478		    	 	& $1.54 \pm 0.65 $ 		&$7.62 \pm 0.32 $ 		&$245 \pm 66 $ 		&$0.55 \pm 0.13 $ 	&$-0.45 \pm 0.42 $	\\	\hline
    \end{tabular}
    \end{center}
\end{table*}
%%%%%%%%%%%%%%%%%%%%%%%%%%%%

We calculate residuals between the observed and the best-fit profiles in each radial bin. The residuals are computed as $(y_{i} - T_{\rm b}(r_{i}))$, where $T_{\rm b}(r_{i})$ is the best-fit profile in each radial bin. Based on these residuals, we also compute the mean absolute relative residual within each set of five radial bins within $r_{\rm cool}$ as well as for all the data points for the outer region (i.e., $r > r_{\rm cool}$) using $|(y - T_{\rm b})/T_{\rm b}|$. The profiles of the residuals and the mean absolute relative residuals are shown in the middle and bottom row of each panel in Figure~\ref{fig:kT_fitting}, respectively. The calculation of residuals will also be applied to the radial profiles of the other components.

We also calculate the ratios of the parameters of the temperature profile for \rxcj, A3112, and A478 to those for A4059, since A4059 has the smallest values for each parameter. These ratios are summarized in Table~\ref{tab:temp_ratio}. 

%%%%%%%%%%%%%%%%%%%%%%%%%%%%
\begin{table*}[htbp]
    \begin{center}
    \caption{
        Ratios of the parameters of the temperature profile for \rxcj, A3112, and A478 to those of A4059.
        }\label{tab:temp_ratio}
    \begin{tabular}{lccc}
    \hline\hline	
    Cluster		& $T_{\rm center} / T_{\rm center, A4059}$    	& $T_{\rm peak} / T_{\rm peak, A4059}$            & $r_{\rm cool} / r_{\rm cool, A4059}$    \\	\hline
    RXCJ1504.1-0248	        & $3.42 \pm 0.42 $ & $2.55 \pm 0.13 $ & $4.21 \pm 0.82 $	\\
    A3112	        			& $1.83 \pm 0.24 $ & $1.15 \pm 0.03 $ & $1.16 \pm 0.16 $	\\
    A478		    	   	& $1.27 \pm 0.55 $ & $1.62 \pm 0.07 $ & $3.18 \pm 0.92 $	\\	\hline
    \end{tabular}
    \end{center}
\end{table*}
%%%%%%%%%%%%%%%%%%%%%%%%%%%%

\subsection{ICM electron number density profiles}
\label{sec:ne_prof}

In the same manner as the analysis of the ICM temperature profiles, we also analyze the radial profiles of the electron number density for the sample. A $\beta$-model is frequently used to model the density profile of the ICM \citep[e.g.,][]{Cavaliere1976, Cavaliere1978, Ettori2000}. For cool-core clusters, a double $\beta$-model is preferred to account for the observed profile, because of the strong excess at the cluster center \citep[e.g.,][]{Pointecouteau2004, Santos2008, Henning2009, Santos2010, Ota2013}. Therefore, to fit the observed number density profile for each cluster, we adopt a double $\beta$-model expressed as
\begin{equation}
    \label{eqt:nD}
    \begin{aligned}
        n(r) &= n_1(r) + n_2(r) \\
             &= n_{0,\,1}\left[1+\left(\frac{r}{r_{\rm c,\,1}}\right)^2\right]^{-\alpha_1} 
              + n_{0,\,2}\left[1+\left(\frac{r}{r_{\rm c,\,2}}\right)^2\right]^{-\alpha_2}, \\
    \end{aligned}
\end{equation}
where $n_{0,\,1}$ and $n_{0,\,2}$ represent the values at the center, respectively, $r_{\rm c,\,1}$ and $r_{\rm c,\,2}$ are the core radii of each $\beta$-model, respectively, and $\alpha_1$ and $\alpha_2$ are the slopes of each $\beta$ model, respectively. 

\begin{figure*}
    \begin{center}
        \includegraphics[width=8.5cm]{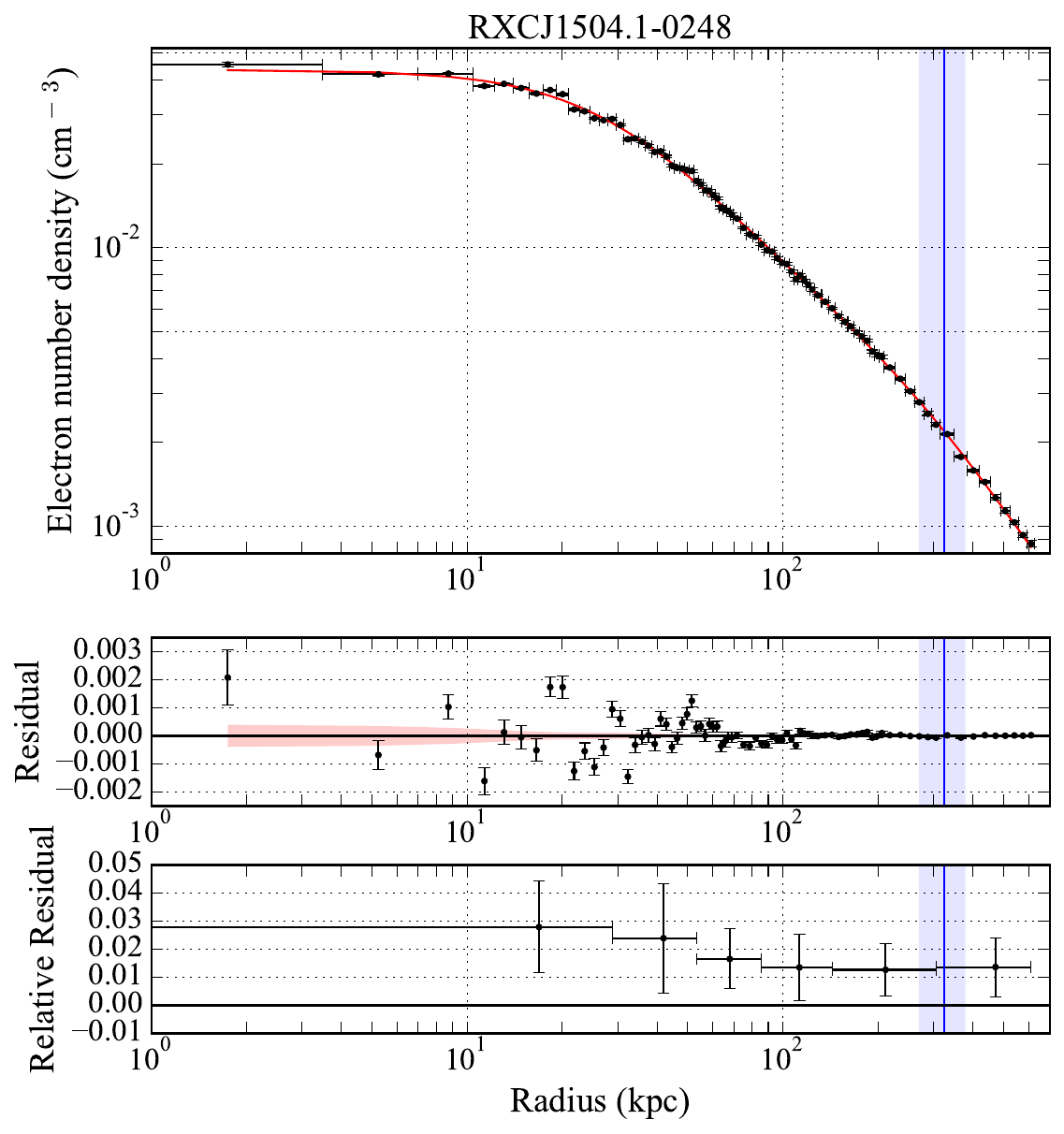}
        \includegraphics[width=8.5cm]{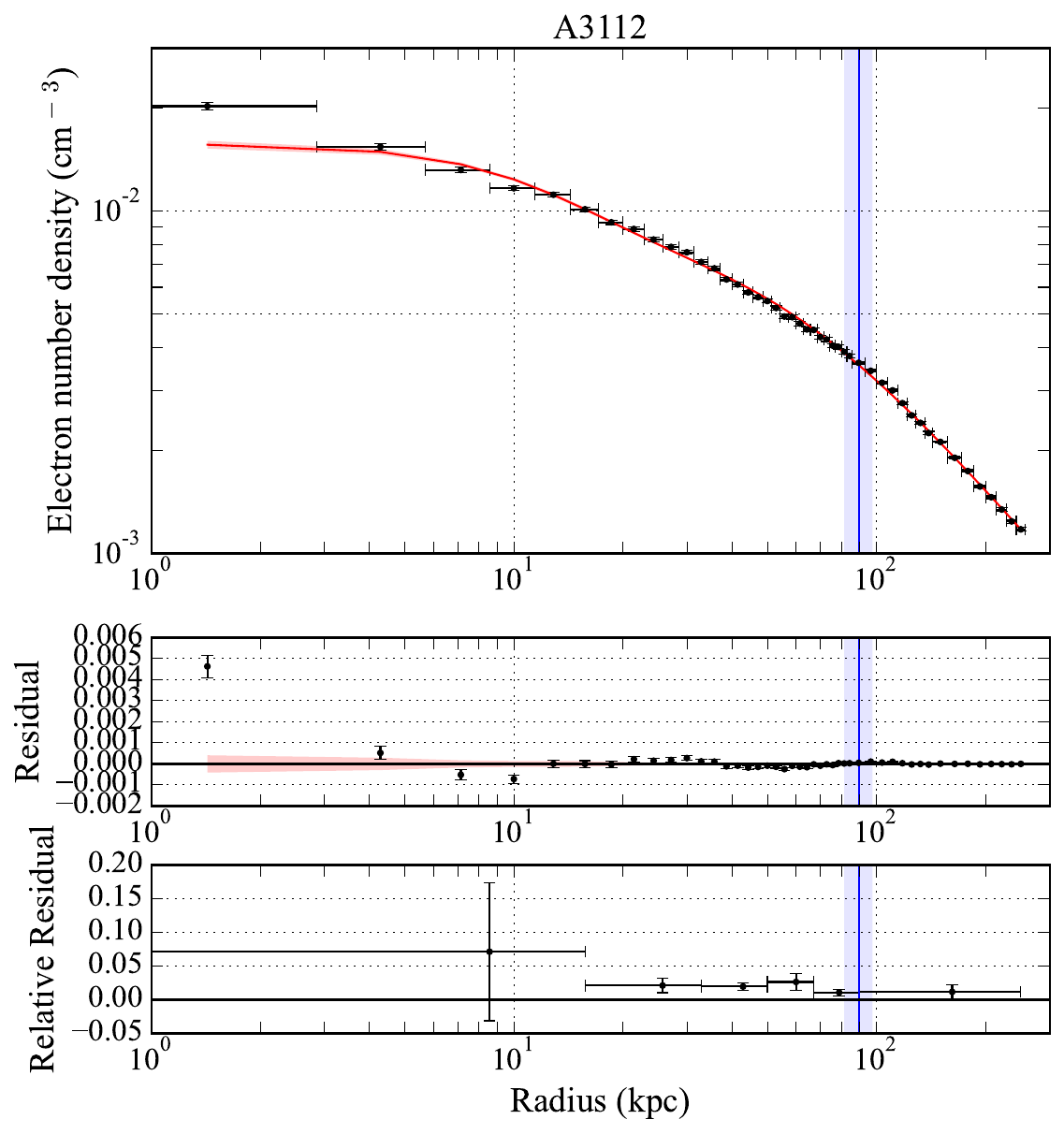}
        \includegraphics[width=8.5cm]{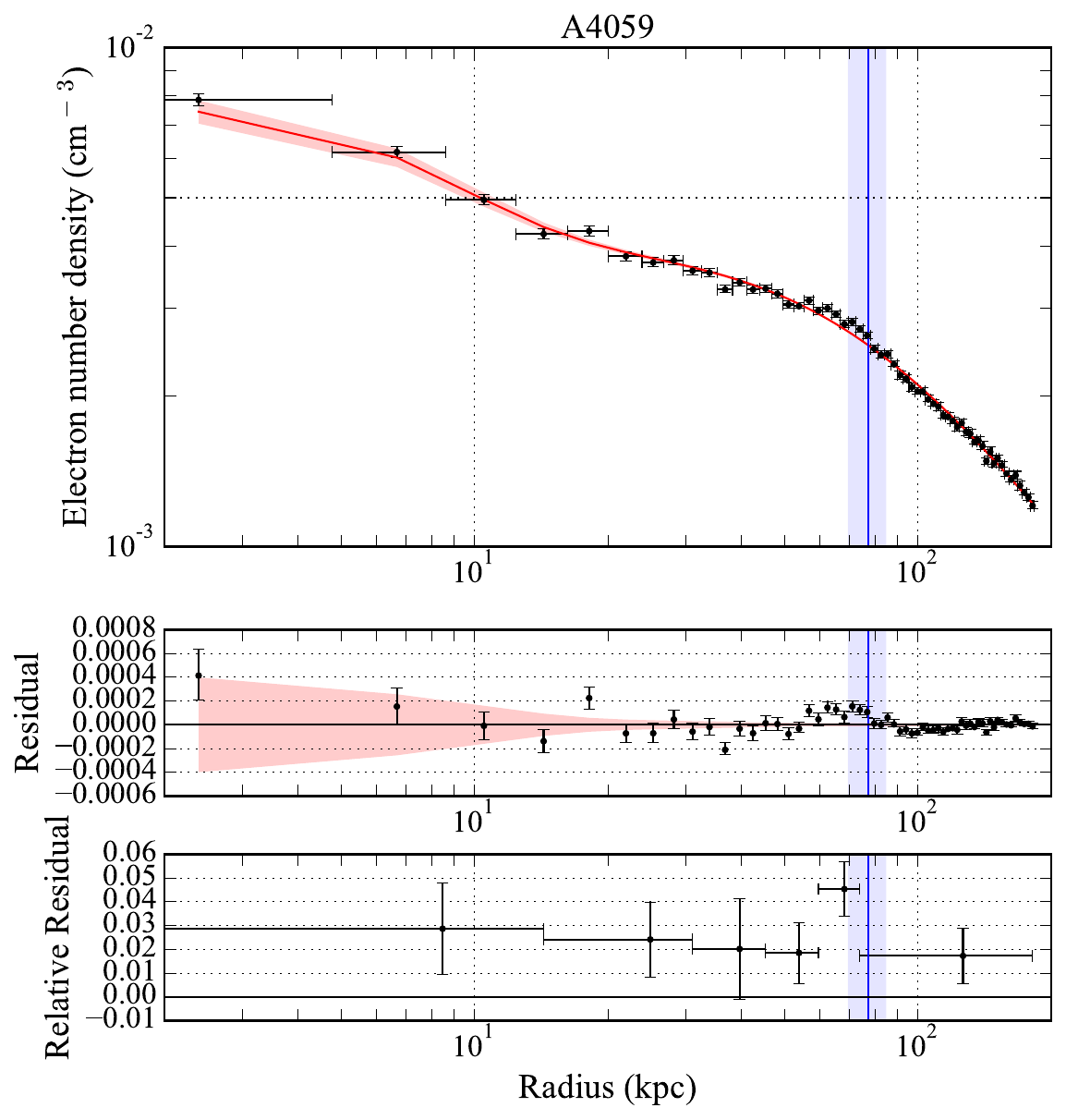}
        \includegraphics[width=8.5cm]{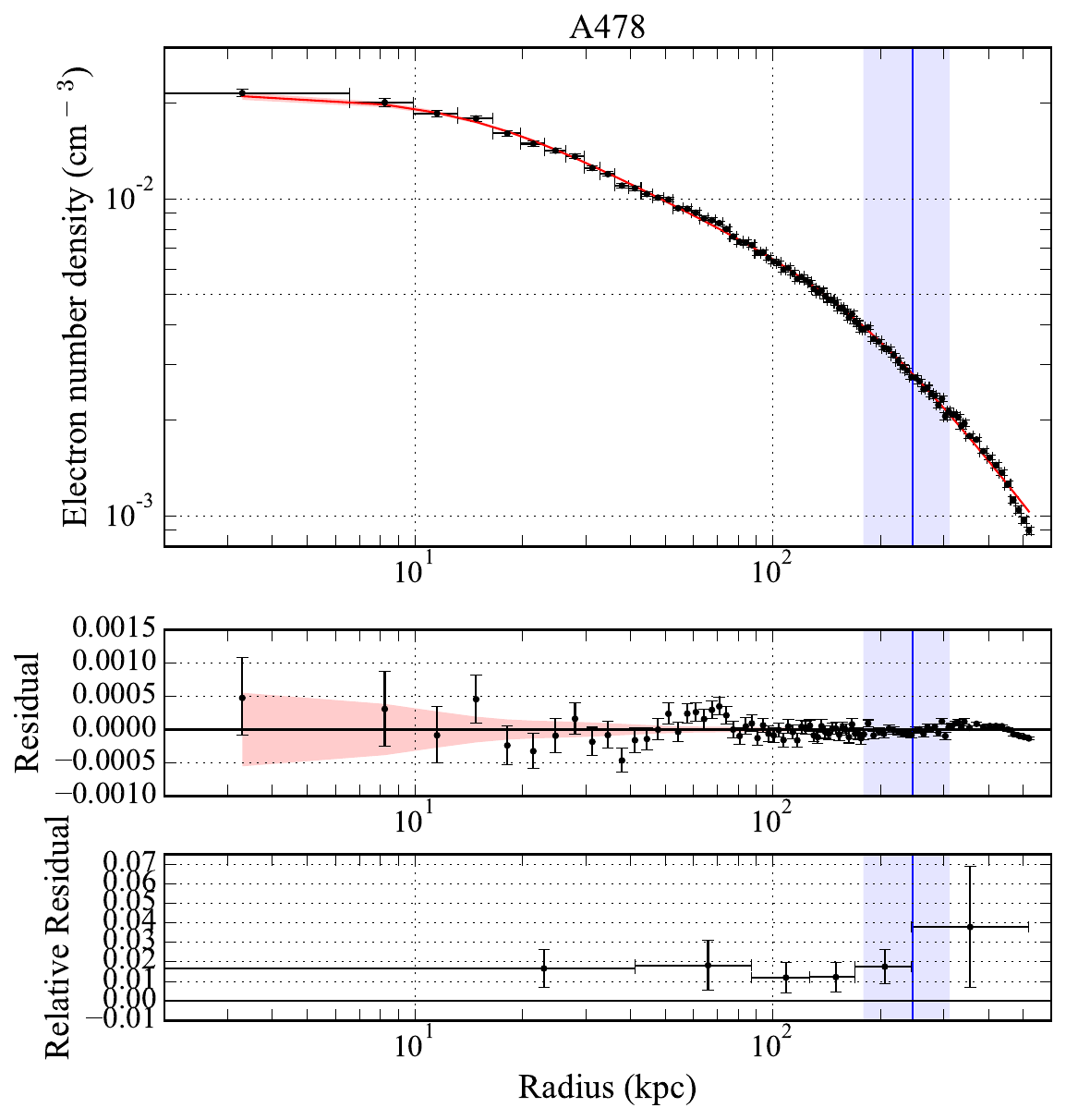}
    \end{center}
\caption{Same as Figure~\ref{fig:kT_fitting} but for the ICM electron number density profiles of the sample: RXCJ1504.1-0248 (top left), A3112 (top right), A4059 (bottom left), and A478 (bottom right).
}
\label{fig:nD_fitting}
\end{figure*}

Following the procedures of the MCMC analysis in Section~\ref{sec:temp_fitting}, we use uninformative uniform priors $n_{0,\,1},\, n_{0,\,2} \in (0, 1)$\,cm\,$^{-3}$, $r_{\rm c,\,1},\, r_{\rm c,\,2} \in (0, 1000)\,{\rm kpc}$, $\alpha_1,\, \alpha_2 \in (0.5, 3.5]$, and ensure $n_{0,\,1} > n_{0,\,2}$. Thus, we fit the observed profile and sample the posterior probability distributions of these six parameters over the full parameter space allowed by the priors. The best-fit parameters obtained from the MCMC analysis are summarized in Table~\ref{tab:nD_best}. The best-fit profile of each cluster along with its uncertainty is shown in Figure~\ref{fig:nD_fitting}.

%%%%%%%%%%%%%%%%%%%%%%%%%%%%
\begin{table*}[htbp]
    \begin{center}
    \caption{
        Parameter constraints of the double $\beta$-model derived from the MCMC analysis.
    }\label{tab:nD_best}
    \begin{tabular}{lcccccc}
    \hline\hline	
    Cluster     	& $n_{\rm 0, 1}$  	    & $r_{\rm c, 1}$        & $\alpha_1$    & $n_{\rm 0, 2}$  	& $r_{\rm c, 2}$        & $\alpha_2$            \\ 
                    & ($10^{-2}$\,cm\,$^{-3}$)&(kpc)                &               & ($10^{-2}$\,cm\,$^{-3}$)      &(kpc)                  &               \\
    \hline
    RXCJ1504.1-0248	& $4.10 \pm 0.15 $ &$30.0 \pm 2.0 $ &$0.72 \pm 0.08 $ &$0.24 \pm 0.12 $ &$329    \pm 111 $ &$1.71 \pm 0.83 $\\
    A3112	        		& $0.80 \pm 0.02 $ &$53.9 \pm 2.6 $ &$0.62 \pm 0.01 $ &$0.78 \pm 0.03 $ &$19.1   \pm 2.9 $ &$2.16 \pm 0.39 $\\
    A4059	       		& $0.40 \pm 0.01 $ &$72.5 \pm 4.6 $ &$0.60 \pm 0.03 $ &$0.38 \pm 0.04 $ &$14.7   \pm 8.8 $ &$2.86 \pm 0.51 $\\
    A478		   	& $1.54 \pm 0.09 $ &$19.8 \pm 2.1 $ &$0.64 \pm 0.10 $ &$0.59 \pm 0.08 $ &$161    \pm 17 $ &$0.85 \pm 0.08 $\\\hline
    \end{tabular}
    \end{center}
    \end{table*}
%%%%%%%%%%%%%%%%%%%%%%%%%%%%

%%%%%%%%%%%%%%%%%%%%%%%%%%%%
%%%%%%%%%%%%%%%%%%%%%%%%%%%%

\subsection{ICM pressure and entropy profiles}
\label{sec:p_K_prof}

\begin{figure*}
    \begin{center}
        \includegraphics[width=8.5cm]{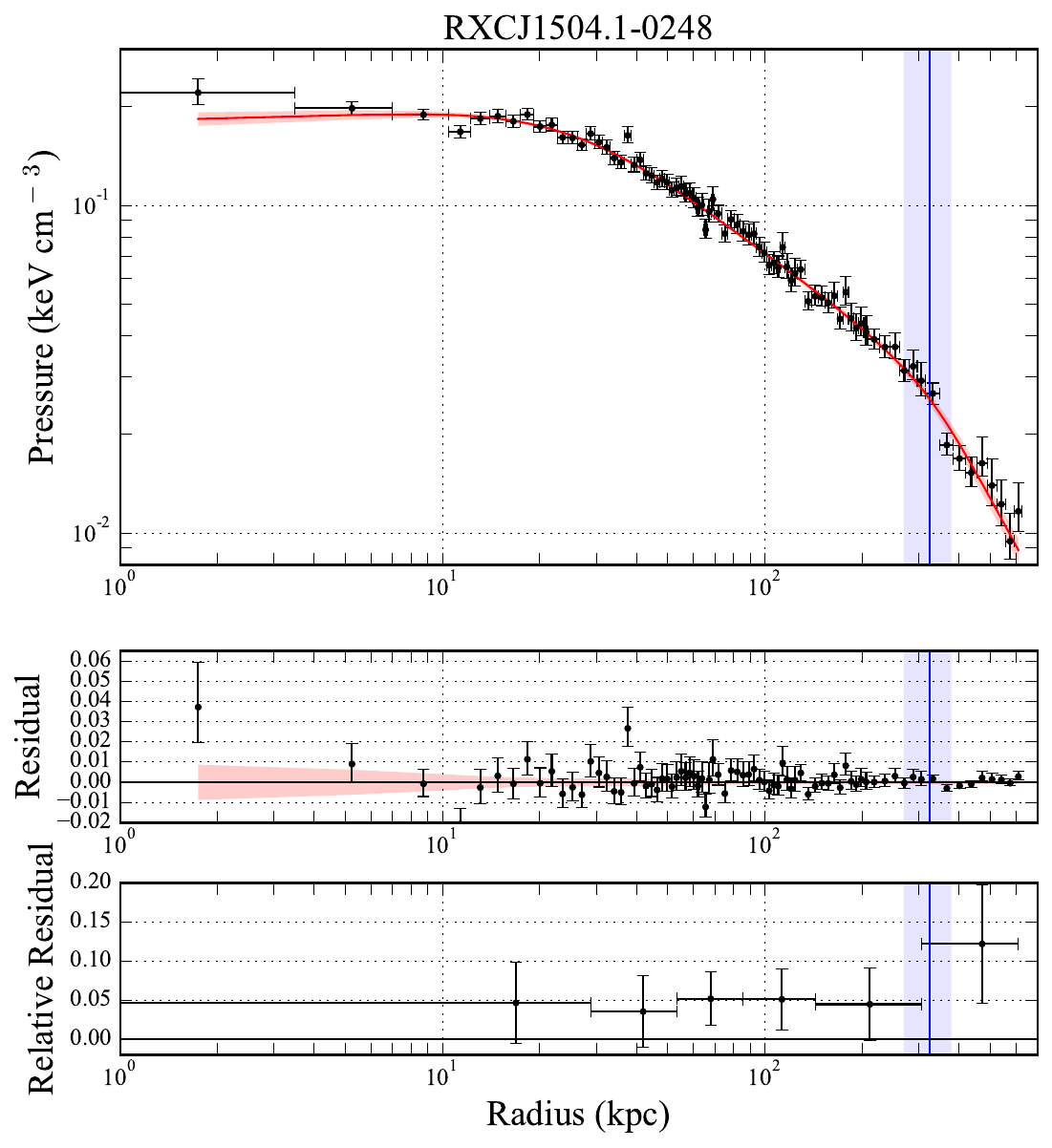}
        \includegraphics[width=8.5cm]{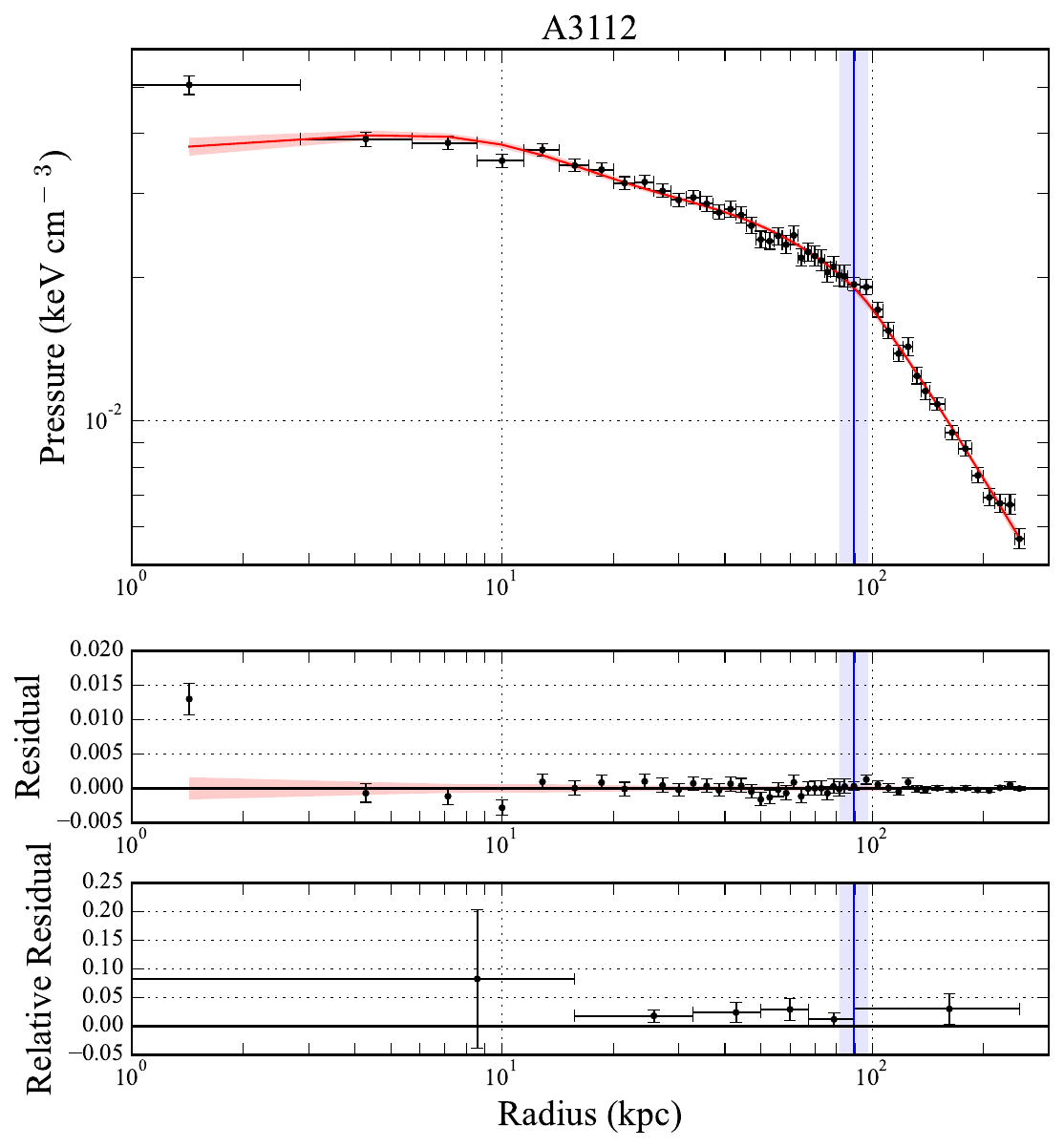}
        \includegraphics[width=8.5cm]{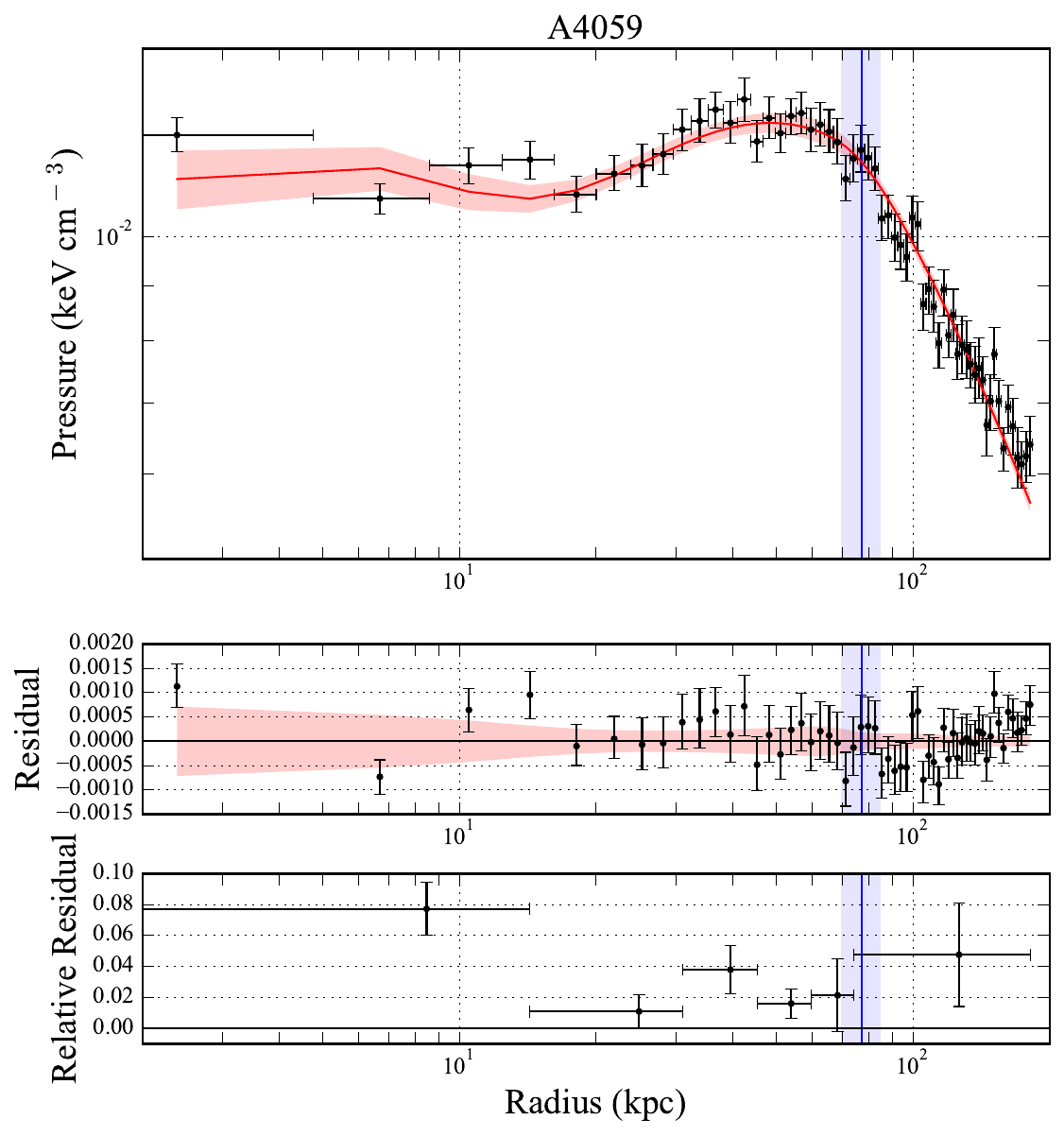}
        \includegraphics[width=8.5cm]{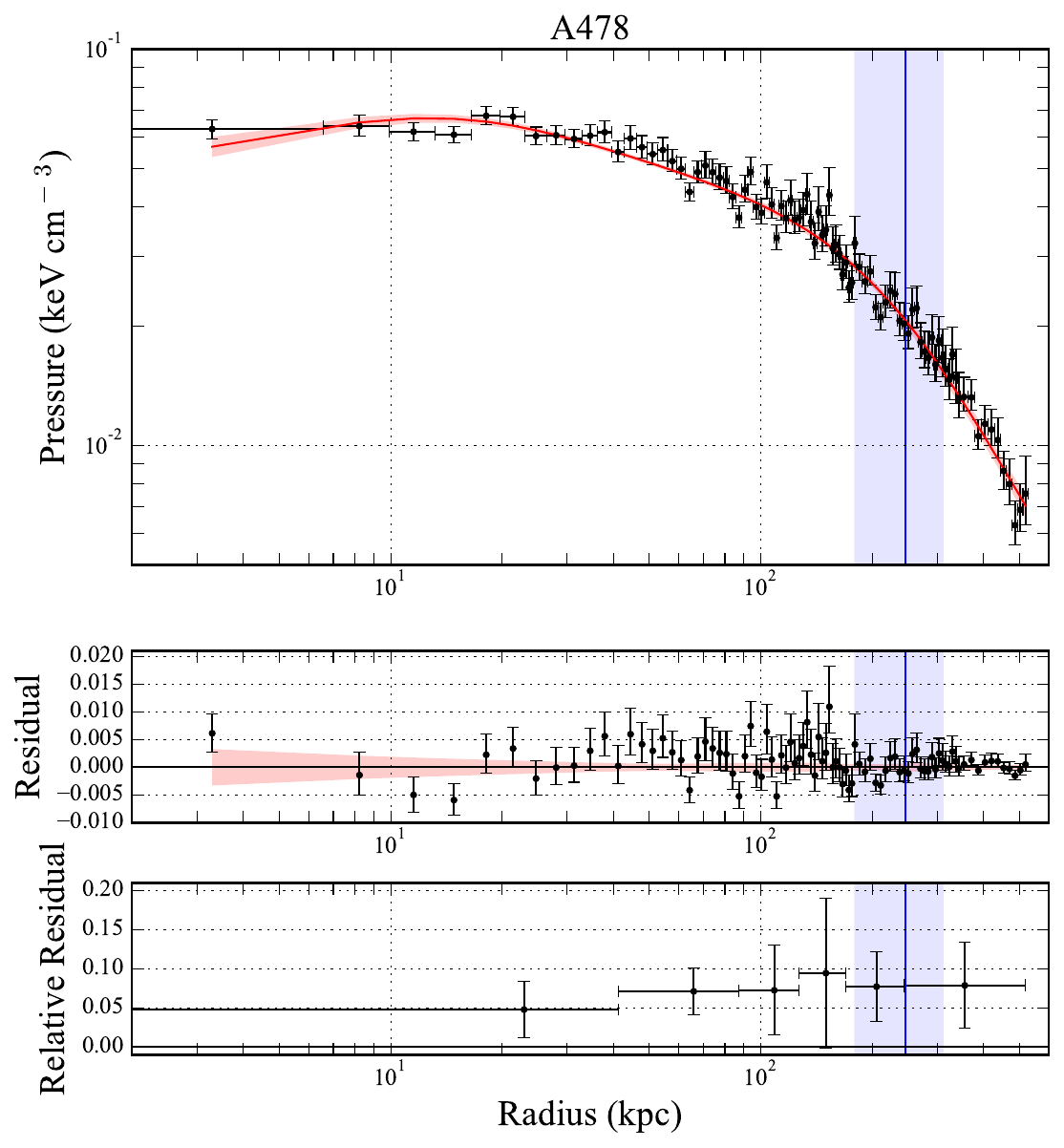}
    \end{center}
\caption{Same as Figure~\ref{fig:nD_fitting} but for the ICM pressure profiles of the sample: RXCJ1504.1-0248 (top left), A3112 (top right), A4059 (bottom left), and A478 (bottom right). 
}
\label{fig:P_fitting}
\end{figure*}

\begin{figure*}
    \begin{center}
        \includegraphics[width=8.5cm]{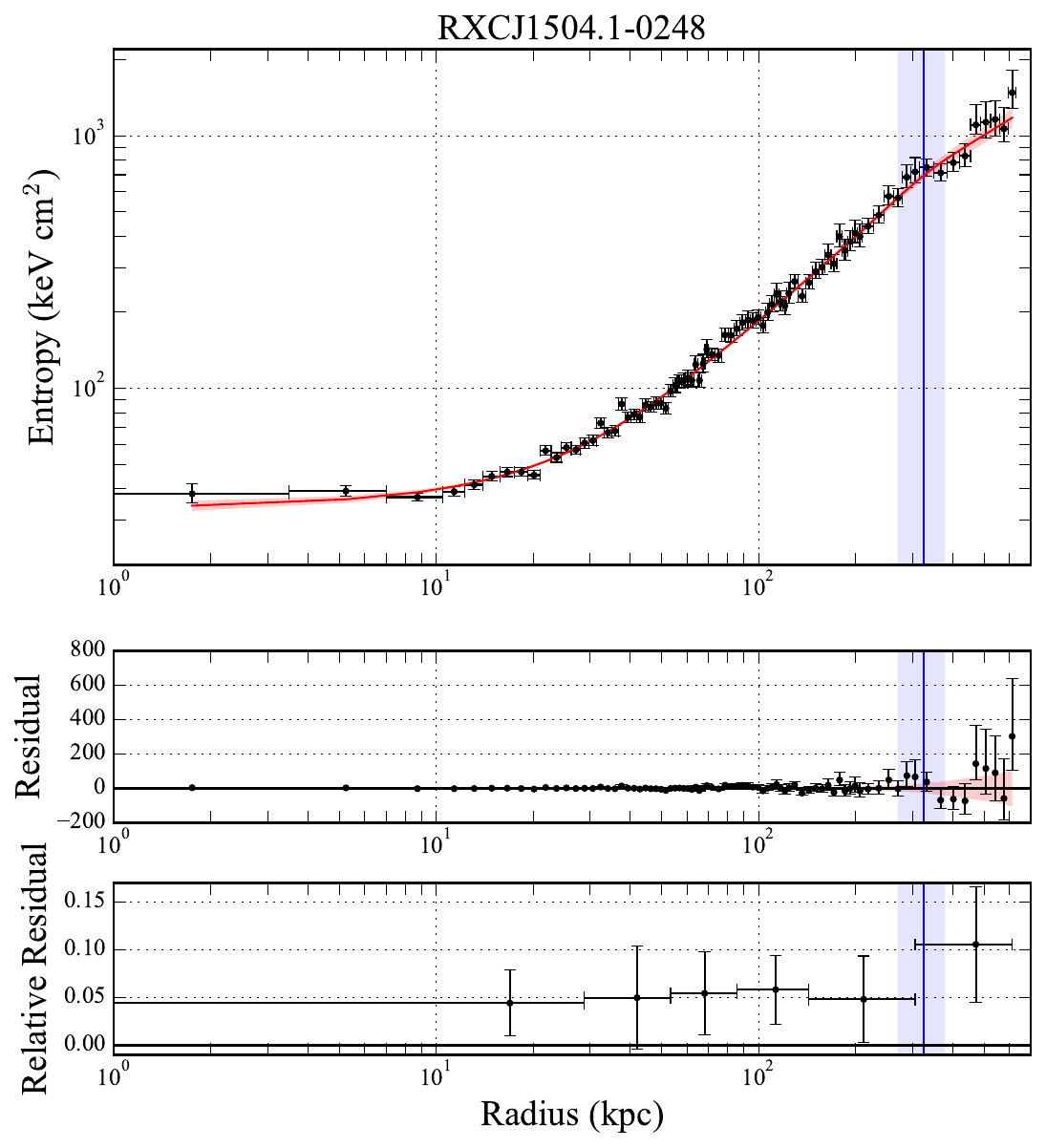}
        \includegraphics[width=8.5cm]{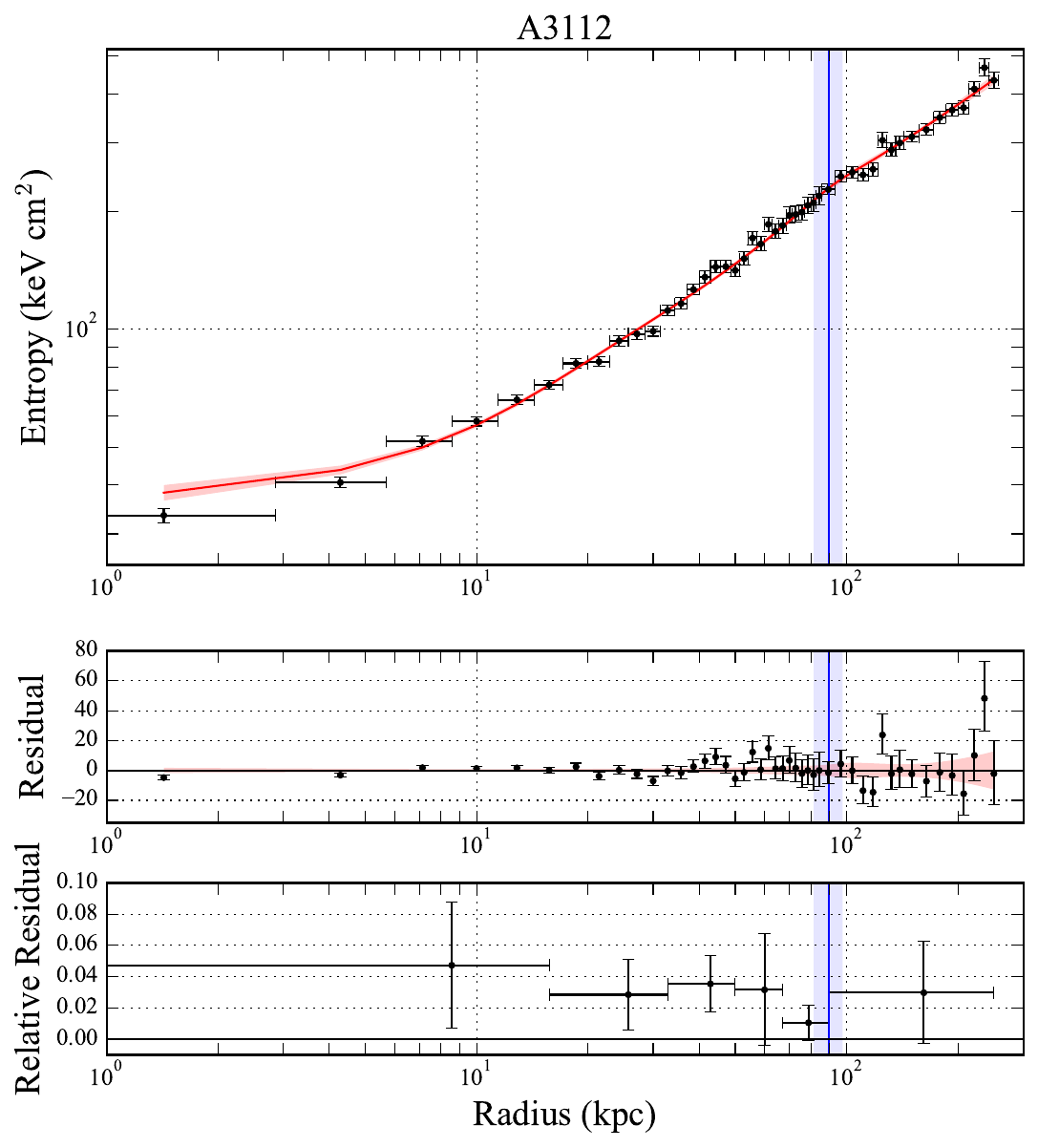}
        \includegraphics[width=8.5cm]{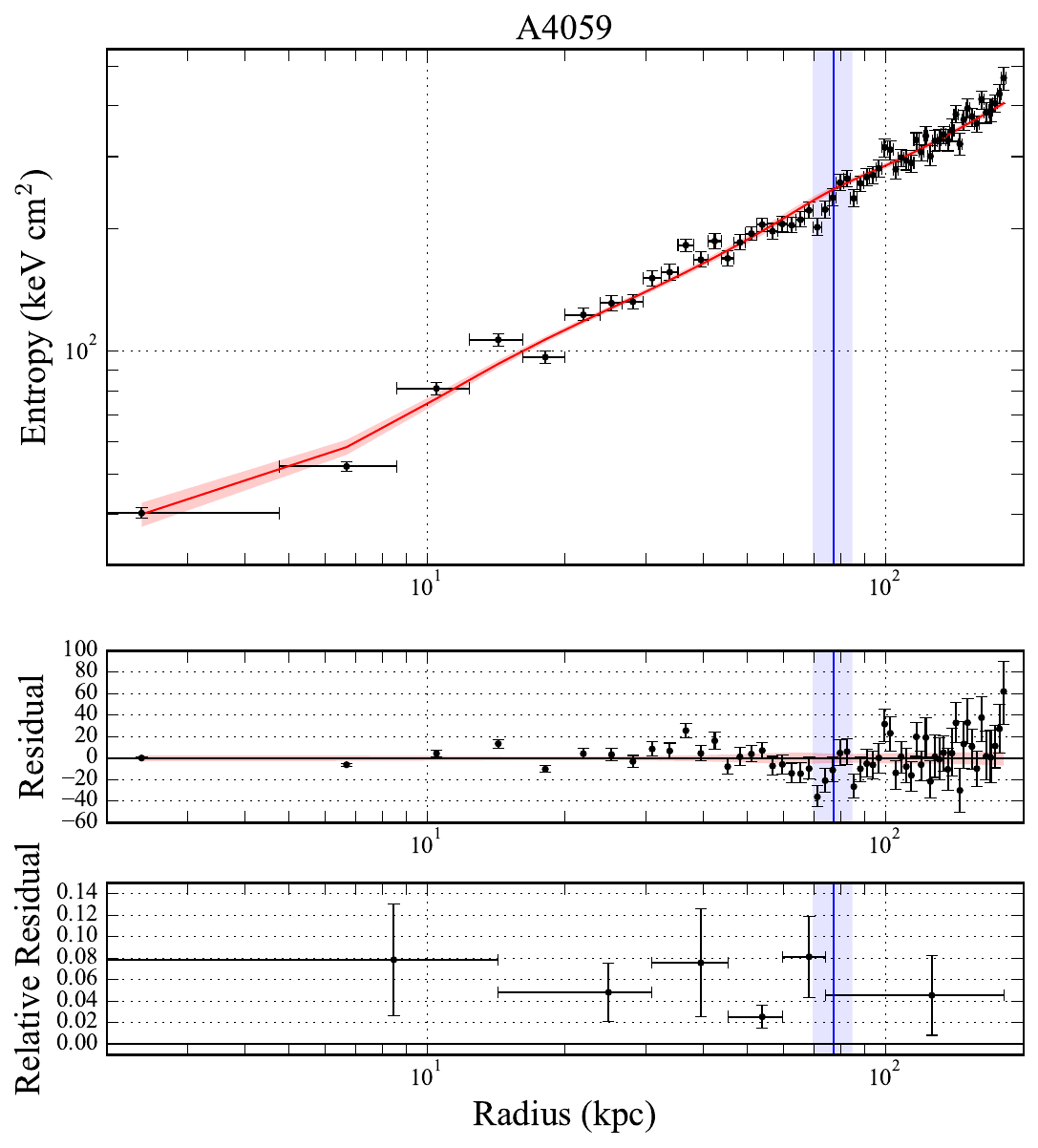}
        \includegraphics[width=8.5cm]{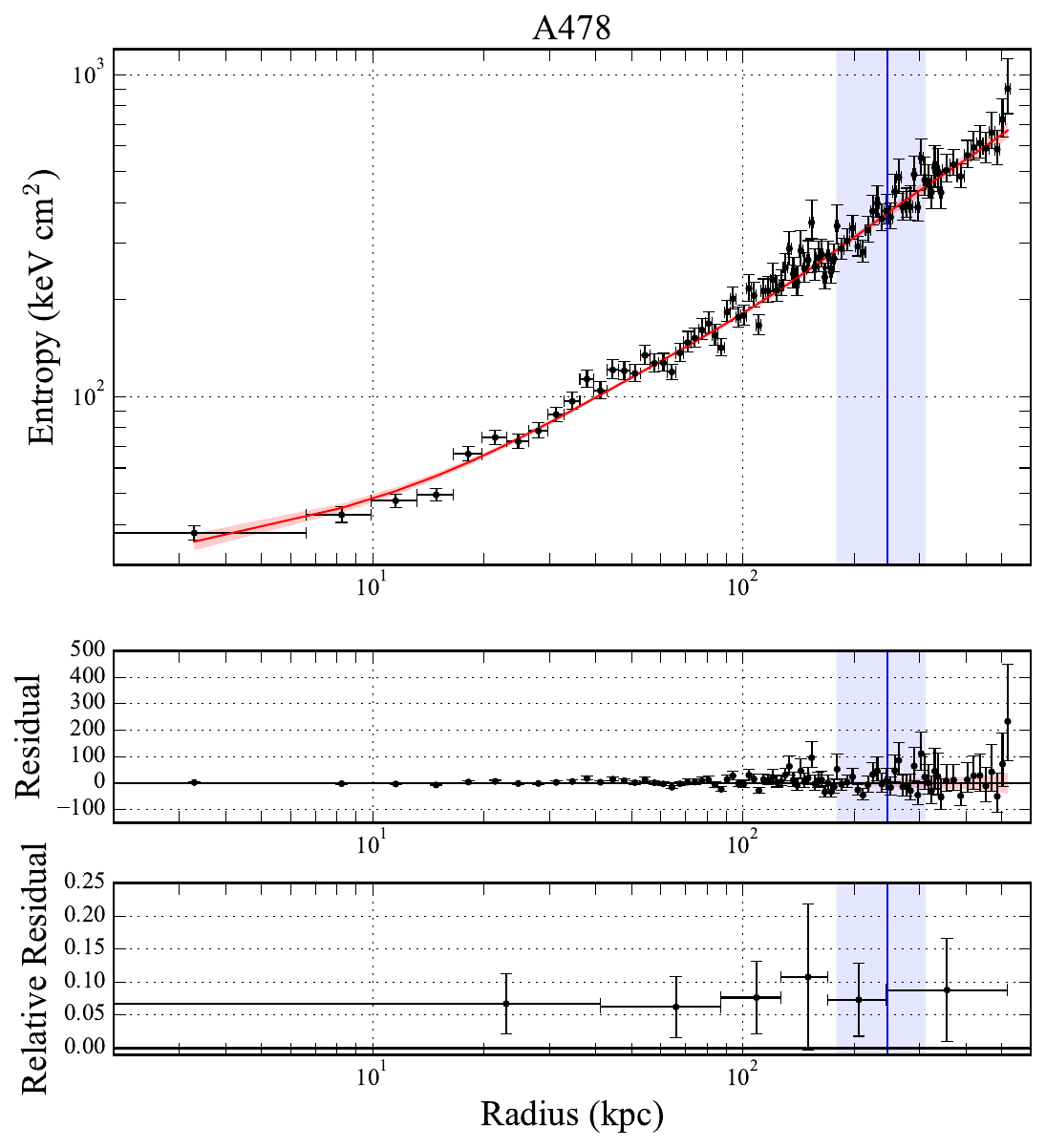}
    \end{center}
\caption{Same as Figure~\ref{fig:nD_fitting} but for the ICM Entropy profiles of the sample: RXCJ1504.1-0248 (top left), A3112 (top right), A4059 (bottom left), and A478 (bottom right).
}
\label{fig:K_fitting}
\end{figure*}

Based on the best-fit profiles for the ICM temperature and electron number density, we compute the predicted profiles for the ICM pressure and entropy profiles using Equations~\ref{eqt:P} and \ref{eqt:K}, respectively. Here, we do not perform direct fitting of the ICM pressure and entropy profiles. The observed ICM pressure and entropy profiles and their predicted profiles are shown in Figures~\ref{fig:P_fitting} and \ref{fig:K_fitting}, respectively.

As mentioned in Section~\ref{sec:A4059}, \cite{Reynolds2008} observed a slight dip in the pressure profile of A4059 at $r \sim 15$\,kpc. We also observe this dip in our pressure profile, which is in agreement with that reported by \cite{Reynolds2008}. The X-ray cavity found by \cite{Reynolds2008} may be associated with this pressure dip and give non-thermal pressure support for this region.

%%%%%%%%%%%%%%%%%%%%%%%%%%%%
%%%%%%%%%%%%%%%%%%%%%%%%%%%%

%%%%%%%%%%%%%%%%%%%%%%%%%%%%
%%%%%%%%%%%%%%%%%%%%%%%%%%%%

%%%%%%%%%%%%%%%%%%%%%%%%%%%%
\begin{figure*}[htbp]
    \begin{center}
     \includegraphics[width=8.5cm]{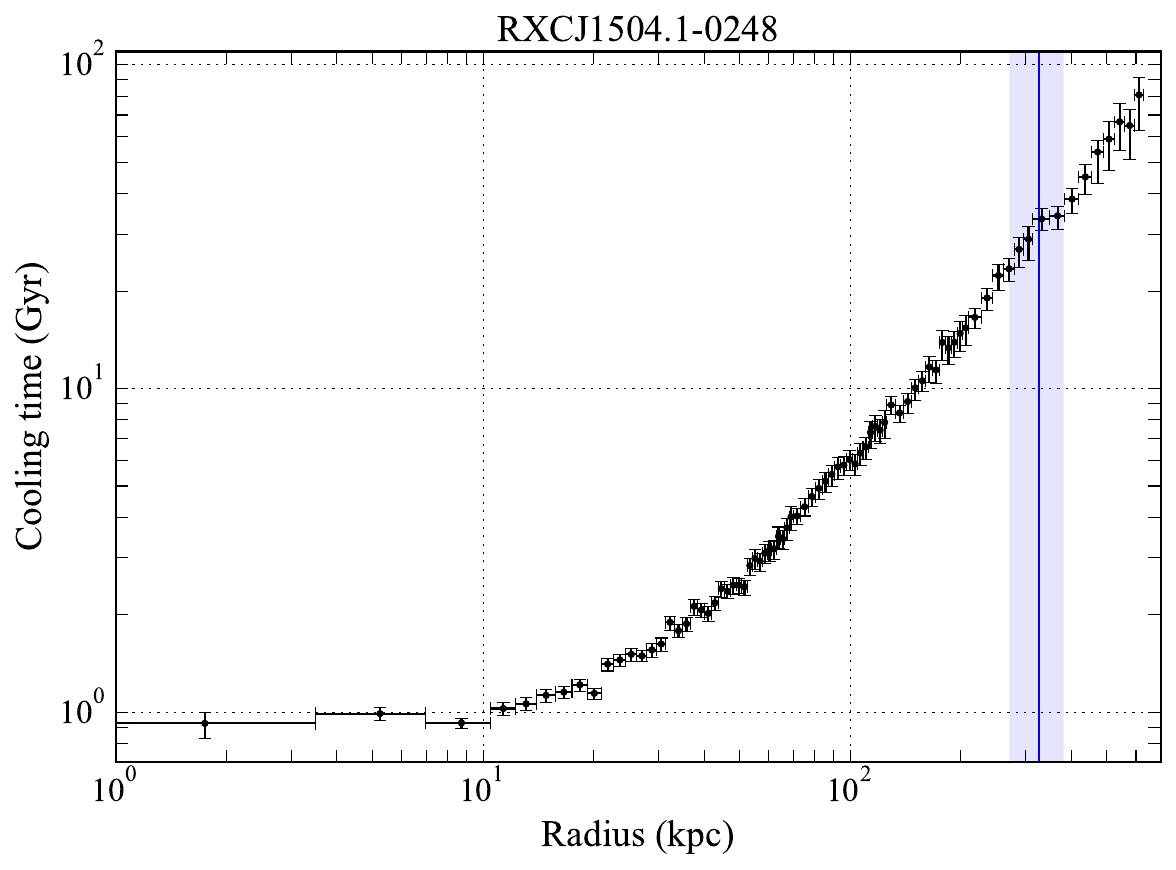}
     \includegraphics[width=8.5cm]{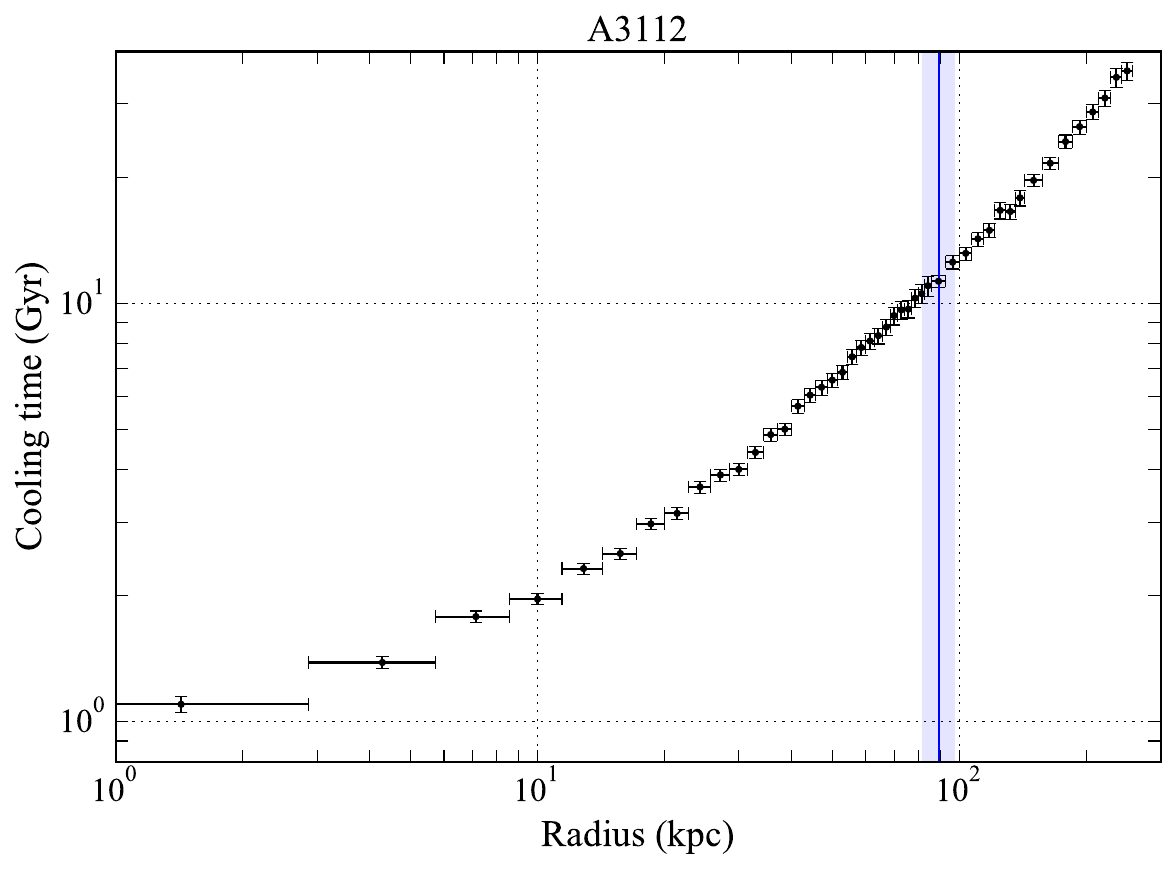}
     \includegraphics[width=8.5cm]{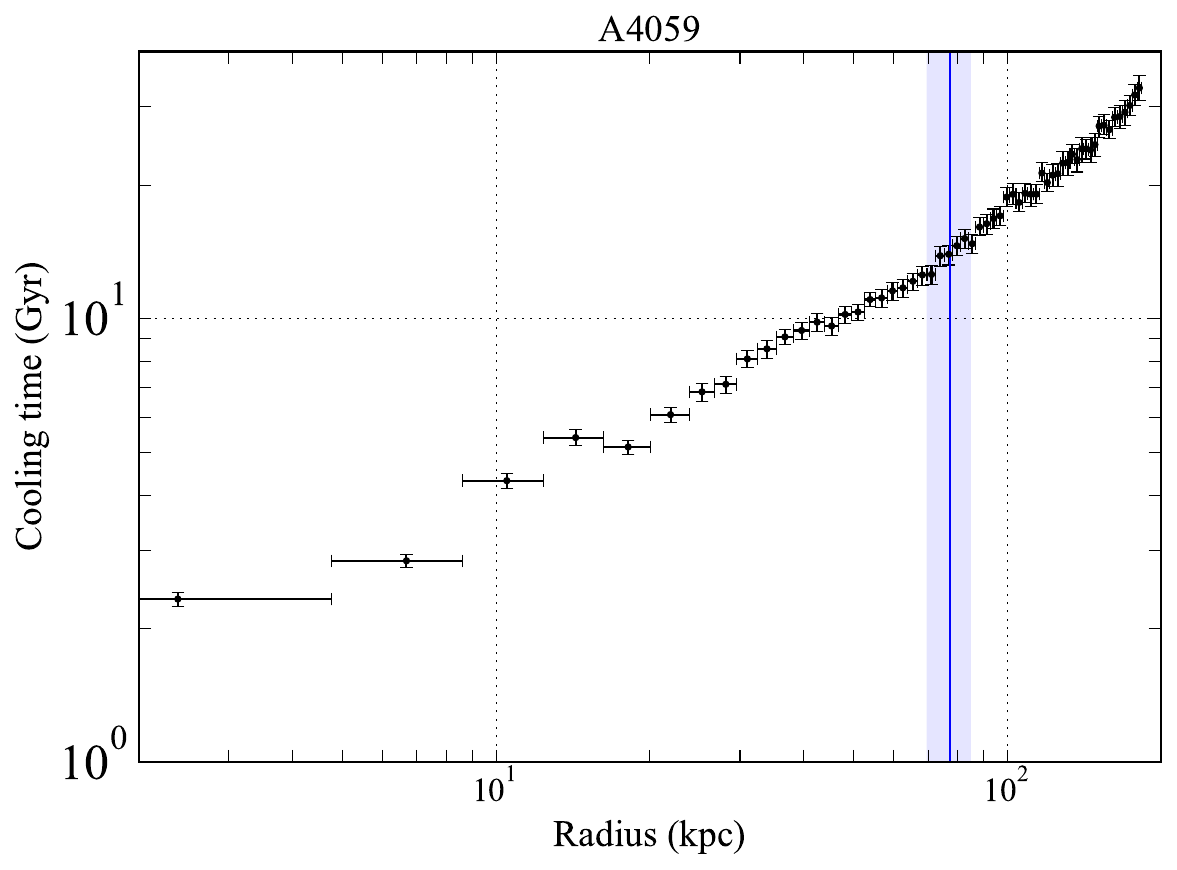}
     \includegraphics[width=8.5cm]{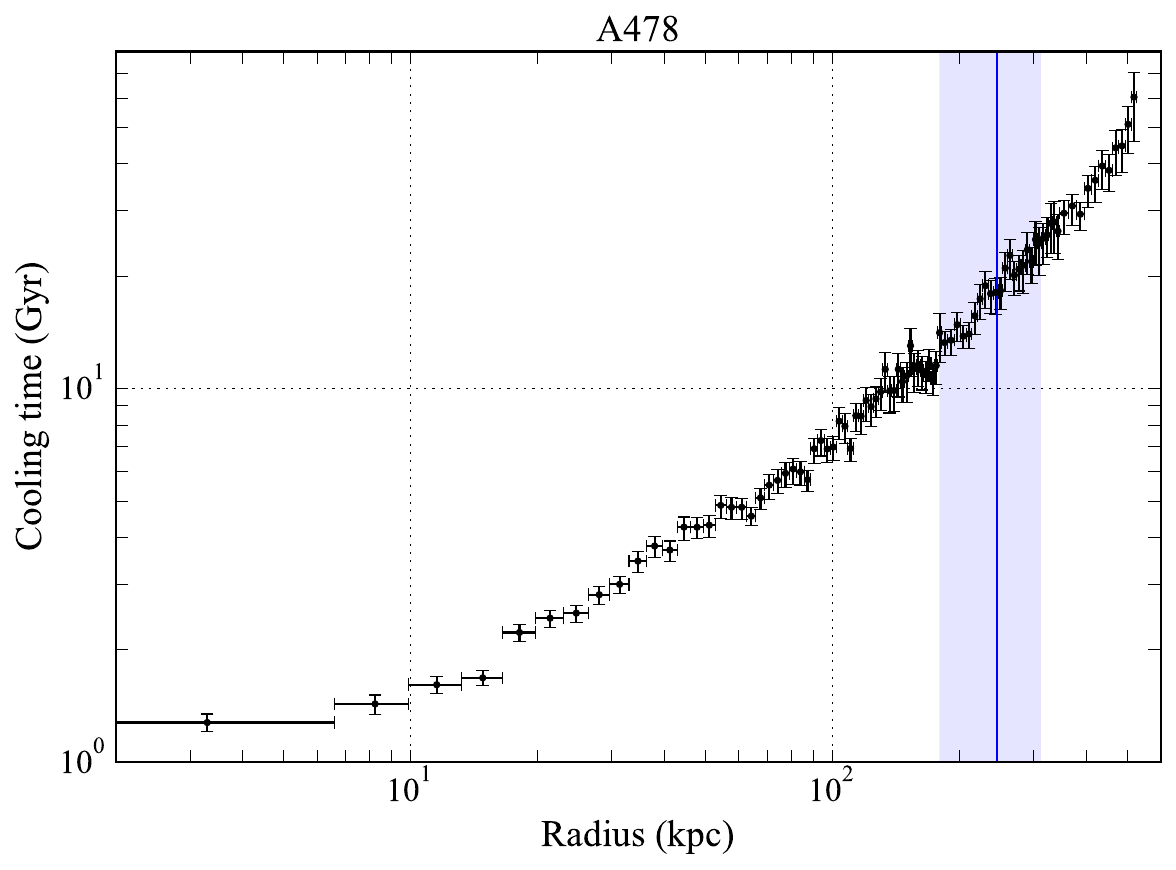}
    \end{center}
\caption{Radial profiles of the ICM radiative cooling time: RXCJ1504.1-0248 (top left), A3112 (top right), A4059 (bottom left), and A478 (bottom right). The blue, vertical solid line and shaded region correspond to the cool-core radius and $1\sigma$ uncertainty for each cluster.
}
\label{fig:tcool_plot}
\end{figure*}
%%%%%%%%%%%%%%%%%%%%%%%%%%%%

\section{Discussion}
\label{sec:discussion}

We have conducted an analysis of the radial profiles of the ICM thermodynamic properties, including temperature, electron number density, pressure, entropy, and radiative cooling time for our sample: \rxcj, A3112, A4059, and A478. We have also determined the cool-core radius for each cluster by analyzing their observed ICM temperature profiles. 

In this section, we discuss the characteristics of the cool-core systems defined by the cool-core radius in the sample. We also explore possible relations between the cool-core radius and the properties of the host galaxy clusters, as well as investigate potential universal forms in the radial profiles scaled by the cool-core radius. Furthermore, we study perturbations in the ICM thermodynamic properties within the cool cores.

\subsection{Cool-core radius}
\label{sec:cc_r}

Cool cores are characterized by a significant drop in the ICM temperature toward the cluster center \citep[][]{Molendi2001}. Since the cool-core radius has been defined as the turnover radius in the ICM temperature profile, the cool-core radius corresponds to a boundary region where the cooling of the ICM becomes dominant. Therefore, the cool-core radius is an important aspect for understanding the underlying physics of cool cores. 

We have determined the cool-core radius for the sample as summarized in Table~\ref{tab:best}. Among our sample, \rxcj ~has the largest cool-core radius ($r_{\rm cool} = 324 \pm 55$\,kpc), while A4059 has the smallest cool-core radius ($r_{\rm cool} = 77 \pm 8$\,kpc). A3112 and A478 exhibit the relatively small and large cool-core radii, respectively. 

We find that the radiative cooling time of the ICM at the cool-core radius exceeds 10\,Gyr for all galaxy clusters in our sample (see Figure~\ref{fig:tcool_plot}). In particular, \rxcj ~exhibits a radiative cooling time of $32^{+5}_{-11}$\,Gyr at its cool-core radius. These time scales are significantly longer than the inferred age of low-$z$ galaxy clusters. Our results indicate that the ICM temperature starts dropping toward the cluster center from a region exhibiting such a long radiative cooling time. Although it is difficult to predict the past thermodynamic properties of the ICM at the same region as the present cool-core radius, it is expected that the effect of radiative cooling in such regions is minimal or negligible. Furthermore, our results show that the radiative cooling time of the ICM gradually increases toward the outskirts, indicating no apparent feature or discontinuity at the cool-core radius. Therefore, our findings indicate that mechanisms for determining the size of cool cores rely not only on radiative cooling but also on additional mechanisms.

Gas sloshing can induce displacement of cool gas originally in a cool core toward the outskirts, leading to mixing of cool gas with ambient hot gas \citep[e.g.,][]{Ascasibar2006, ZuHone2010, Keshet2012, Naor2020, Keshet2023}. Since the cool-core systems in the sample have developed well as indicated by the observed temperature drop, radiative cooling is expected to play a dominant role in generating cool gas in the inner regions of the cool cores. Such cool gas is likely to be displaced toward the outer regions by gas sloshing. In fact, evidence of gas sloshing has been observed in the cool cores in our sample \citep[][]{Ueda2021}, supporting this hypothesis.

\subsection{Relation between the cool-core radius and the cluster mass}
\label{sec:rcc_vs_r500}

To explore possible relations between the cool-core radius and the cluster mass, we extract the $M_{500}$ and $r_{500}$ values for our sample from the MCXC catalog \citep[][]{Piffaretti2011}. These values are summarized in Table~\ref{tab:mass}. Since $r_{500}$ is commonly used as a scaling factor for the radial profiles of the ICM thermodynamic properties, this study provides insights into the connection between the cool-core radius and $r_{500}$.

%%%%%%%%%%%%%%%%%%%%%%%%%%%%
\begin{table*}[htb!]
    \begin{center}
    \caption{
        Cool-core radius, cluster mass $M_{500}$, fiducial radius $r_{500}$, and those scaled by the values of A4059.
        } \label{tab:mass}
    \begin{tabular}{lcccccccc}
    \hline\hline	
    Cluster		&  $r_{\rm cool}$       &$r_{\rm cool} / r_{\rm cool, A4059}$     & $ r_{\rm 500} $ & $r_{\rm cool}/r_{500}$	& $ r_{\rm 500} / r_{\rm 500, A4059} $  & $M_{\rm 500} $        & $ M_{\rm 500} / M_{\rm 500, A4059} $ 	 \\
                & (kpc)	                    & 			                              & (Mpc)			&	&		                            & ($10^{14}$\,\MO) &						\\ 
    			\hline
    RXCJ1504.1-0248	        &$325 \pm 55$ 		& $4.21 \pm 0.82 $ 	& 1.52 & $0.21 \pm 0.04$		& 1.58 & 12.47 & 4.68\\
    A3112	        			&$89.4 \pm 8.0$	& $1.16 \pm 0.16 $ 	& 1.13 & $0.080 \pm 0.007$	&1.17 & 4.39 & 1.65\\
    A4059	        			&$77.1 \pm 7.7 $    	& 1.00          		& 0.96 & $0.080 \pm 0.008$	&1.00 & 2.67 & 1.00\\
    A478	        			&$245 \pm 66 $  	& $3.18 \pm 0.92 $ 	& 1.28 & $0.19 \pm 0.05$		&1.32 & 6.42 & 2.41\\ \hline
    \end{tabular}
    \end{center}
\end{table*}
%%%%%%%%%%%%%%%%%%%%%%%%%%%%

To investigate the relations of the cool-core radius to $M_{500}$ and to $r_{500}$, we conduct a regression analysis using a model expressed as
\begin{equation}
    \label{eqt:r500M500}
    \frac{r_{\rm cool}}{\hat{r}_{\rm cool}} = N_i \left(\frac{x_{i}}{\hat{x}_i} \right)^{\alpha_i},
\end{equation}
where $N_i$ is a normalization of the model defined as $N_i = 10^{A_i}$, $x_i$ represents either $M_{500}$ or $r_{500}$, and $\alpha_i$ is the slope of the model. Here, $i$ denotes the values for $M_{500}$ or $r_{500}$. We center the relation on the pivot values $\hat{r}_{\rm cool} = 167$\,kpc, $\hat{M}_{500} = 5.41 \times 10^{14}$\,\MO, and $\hat{r}_{500} = 1.20$\,Mpc set at the median of the distributions of $r_{\rm cool}$, $M_{500}$, and $r_{500}$, respectively. 

\begin{figure*}
    \begin{center}
        \includegraphics[width=8.5cm]{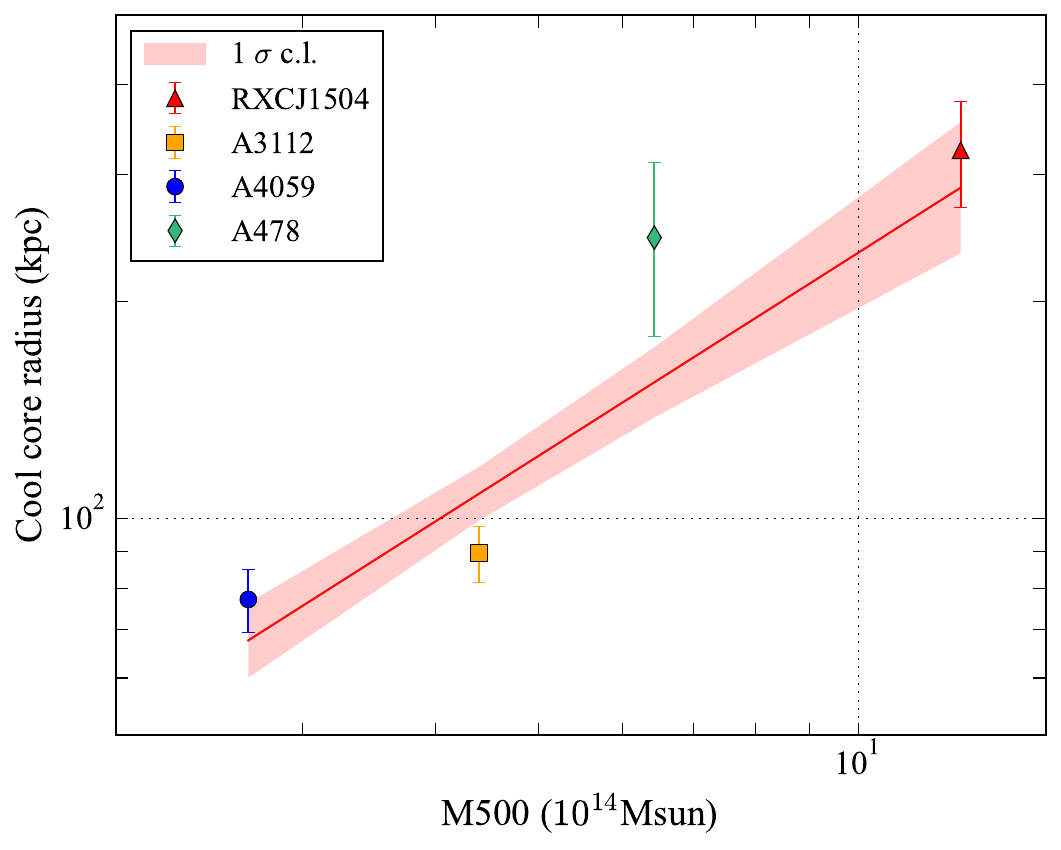}
        \includegraphics[width=8.5cm]{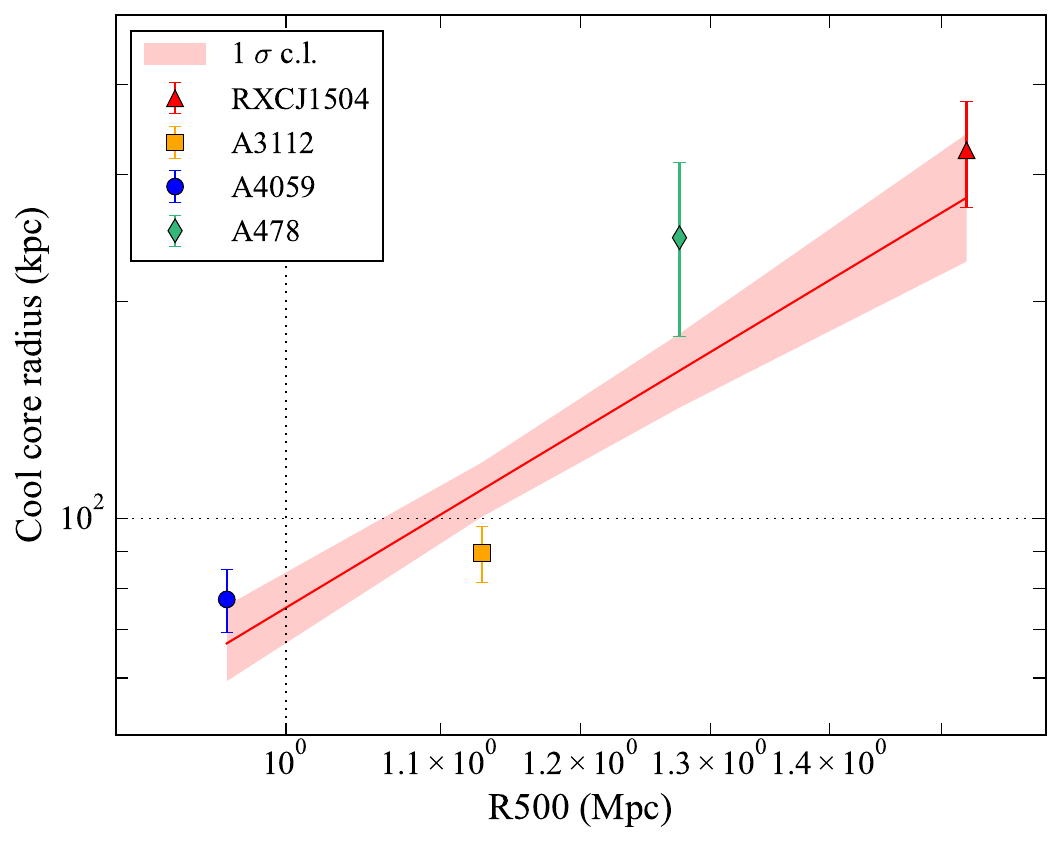}
    \end{center}
\caption{Relations of the cool-core radius ($r_{\rm cool}$) to $M_{500}$ and to $r_{500}$. 
Left: the relation between $r_{\rm cool}$ and $M_{500}$. The red solid line and shaded region show the best-fit relation obtained from the fitting in the log-log space and its 1\,$\sigma$ uncertainty, respectively.
Right: Same as the left panel but for the relation between $r_{\rm cool}$ and $r_{500}$.
}
\label{fig:r500M500}
\end{figure*}

Following the procedures of the MCMC analysis in Section~\ref{sec:temp_fitting}, we use uninformative uniform priors $A_r,\, A_M \in (-5, 5)$ and $\alpha_r,\, \alpha_M \in (0, 10)$. We fit the data in the log-log space and sample the posterior probability distributions of the parameters over the full parameter space allowed by the priors. The best-fit parameters for the relations of $r_{\rm cool}$ to $M_{500}$ and to $r_{500}$ are summarized in Table~\ref{tab:r500M500}. The best-fit relations for $M_{500}$ and $r_{500}$ with $1\sigma$ uncertainty are shown in Figure~\ref{fig:r500M500}.

%%%%%%%%%%%%%%%%%%%%%%%%%%%%
\begin{table*}[htbp]
    \begin{center}
    \caption{
        Best-fit parameters of the relation of $r_{\rm cool}$ to $M_{500}$ and to $r_{500}$.
        }\label{tab:r500M500}
    \begin{tabular}{ccc|ccc}
    \hline\hline	
    \multicolumn{3}{c|}{$r_{\rm cool}$ vs. $M_{500}$} 	& \multicolumn{3}{c}{$r_{\rm cool}$ vs. $r_{500}$}            \\ \hline
    $A_M$&$N_M= 10^{A_M}$ & $\alpha_M$ & $A_r$	&$N_r= 10^{A_r}$ & $\alpha_r$              \\
    \hline
    $-0.10 \pm 0.04 $ &$0.79_{-0.07}^{+0.08} $ &$0.94 \pm 0.18 $ & $-0.10 \pm 0.04$ & $0.80_{-0.07}^{+0.08} $ &$3.12 \pm 0.59 $	\\\hline
    \end{tabular}
    \end{center}
    \end{table*}
%%%%%%%%%%%%%%%%%%%%%%%%%%%%
    
%%%%%%%%%%%%%%%%%%%%%%%%%%%%
%%%%%%%%%%%%%%%%%%%%%%%%%%%%

Based on the analysis with a very limited sample size (i.e., only four clusters), we find that the best-fit relation of the cool-core radius to $M_{500}$ is consistent with a linear relation (i.e., $\alpha_M = 0.94 \pm 0.18$). Since $M_{500}$ is proportional to $r_{500}^3$, the best-fit slope for the $r_{\rm cool}$-$r_{500}$ relation is also consistent with this dependency (i.e., $\alpha_r = 3.12 \pm 0.59$). However, it is difficult to arrive at a robust conclusion regarding these relations due to such a very small sample size. Further study should be conducted to interpret these results robustly. The detailed study of these relations with a larger sample size will be reported in a separate paper (Ng \& Ueda 2024, in preparation).

Despite such limitations, here, we attempt to investigate the possible mechanism(s) for the observed relations. If we assume $r_{\rm cool} \propto L_{\rm X}$, meaning that the size of cool cores depends on the X-ray luminosity of their host galaxy clusters, can we infer a slope for the $r_{\rm cool}$-$M_{500}$ relation? Since the cluster mass $M_{500}$ was estimated using the luminosity-mass scaling relation from \cite{Arnaud2010} ($L_{\rm X} \propto M^{1.64}$), with the X-ray luminosity in the $0.1 - 2.4$\,keV band within $r_{500}$ being used in the calculation \citep[see][]{Piffaretti2011}, a slope of $\sim 1.6$ may be expected for the $r_{\rm cool}$-$M_{500}$ relation. Previous studies found that the observed slope for the luminosity-mass scaling relation falls within the range of $1.3 \sim 2$ \citep[see a review][and references therein]{Giodini2013}. Furthermore, the self-similar model predicts a slope of $4/3$ for the luminosity-mass relation. On the other hand, if $r_{\rm cool} \propto t_{\rm cool}$, $t_{\rm cool} \propto n^{-1} \sqrt{T}$ may yield a slope of $1/3$ for the $r_{\rm cool}$-$M_{500}$ relation. However, the best-fit slope is inconsistent with these slopes.

The observed linear relation between $r_{\rm cool}$ and $M_{500}$ indicates that not only baryon physics such as cooling, heating, and heat transport is crucial for the formation of cool cores, but also the size of cool cores is linked to the evolution of their host galaxy clusters. Once the mass of a galaxy cluster increases owing to a cluster merger, the ICM density at the center is expected to increase after the relaxation of merger events. Then, radiative cooling becomes more efficient, leading to the accumulation of a large amount of cool gas in the central region. Since galaxy clusters grow through continuous accretion of material, including cluster mergers, from their surrounding large-scale environments, gas sloshing is expected to take place in cool cores continuously. In fact, this hypothesis is supported by recent comprehensive studies of cool cores \citep[][]{Ueda2020, Ueda2021}. Therefore, the continuous occurrence of gas sloshing may play a crucial role not only in determining the size of cool cores but also in suppressing runaway cooling of the ICM. 

AGN feedback is also expected to play a crucial role not only in transporting gas from the central regions to the outer regions but also in contributing significantly to heating the ICM in cool cores. X-ray cavities are commonly observed in cool cores and considered to be generated by AGN activities such as radio jets \citep[e.g.,][]{Birzan2004, Rafferty2006, Hlavacek-Larrondo2015, Shin2016}. In fact, the cavity power correlates with the X-ray luminosity in the corresponding regions. Numerical simulations of AGN feedback indicate that the self-regulated feedback of AGN is one of the promising heating sources to explain the observed stability of cool cores \citep[e.g.,][]{Gaspari2011, Yang2016}. Hence, the size of cool cores may also be influenced by AGN feedback. Cool cores may coevolve with their host galaxy clusters.

For the relation between the cool-core radius and a fiducial radius, \cite{Vikhlinin2005} reported that the projected temperature as a function of radius reaches a maximum at $r \sim 0.1 - 0.2\,r_{180}$. \cite{Rasmussen2007} studied the relation between the peak position in the temperature profile (i.e., $r_{\rm cool}$) and $r_{500}$ using a sample of 15 nearby galaxy groups observed with \Chandra. Assuming a linear relation, they fitted the data and obtained $r_{\rm cool} = (0.20 \pm 0.02)\,(r_{500}/{\rm kpc}) - (46.7 \pm 15.1)\,{\rm kpc}$. \cite{OSullivan2017} also studied the relation between the turnover radius and $r_{500}$ using a sample of high-richness local galaxy groups, finding that the observed relation in their sample is mostly consistent with that presented by \cite{Rasmussen2007}. In addition, \cite{Rasmussen2007} and \cite{OSullivan2017} pointed out that the relation for galaxy groups differs from that observed in a sample of local galaxy clusters \citep{Vikhlinin2005}, suggesting that this discrepancy may be attributed to differences in the physical properties between galaxy clusters and groups. In fact, our best-fit relation between $r_{\rm cool}$ and $r_{500}$ seems to deviate from that observed for galaxy groups. Since galaxy groups have a lower temperature gas compared to galaxy clusters, radiative cooling becomes more efficient owing to emission lines from the metals in the ICM. Additionally, mergers may have a more significant impact on a cool core of the primary galaxy group. Perhaps, the $r_{\rm cool}$-$r_{500}$ relation may have a break. Further study is required to arrive at a firm conclusion regarding the similarities and differences for the $r_{\rm cool}$-$r_{500}$ relation between galaxy clusters and groups.

\subsection{Scaled radial profiles of the ICM thermodynamic properties}
\label{sec:scaled}

\begin{figure}
    \begin{center}
        \includegraphics[width=8.5cm]{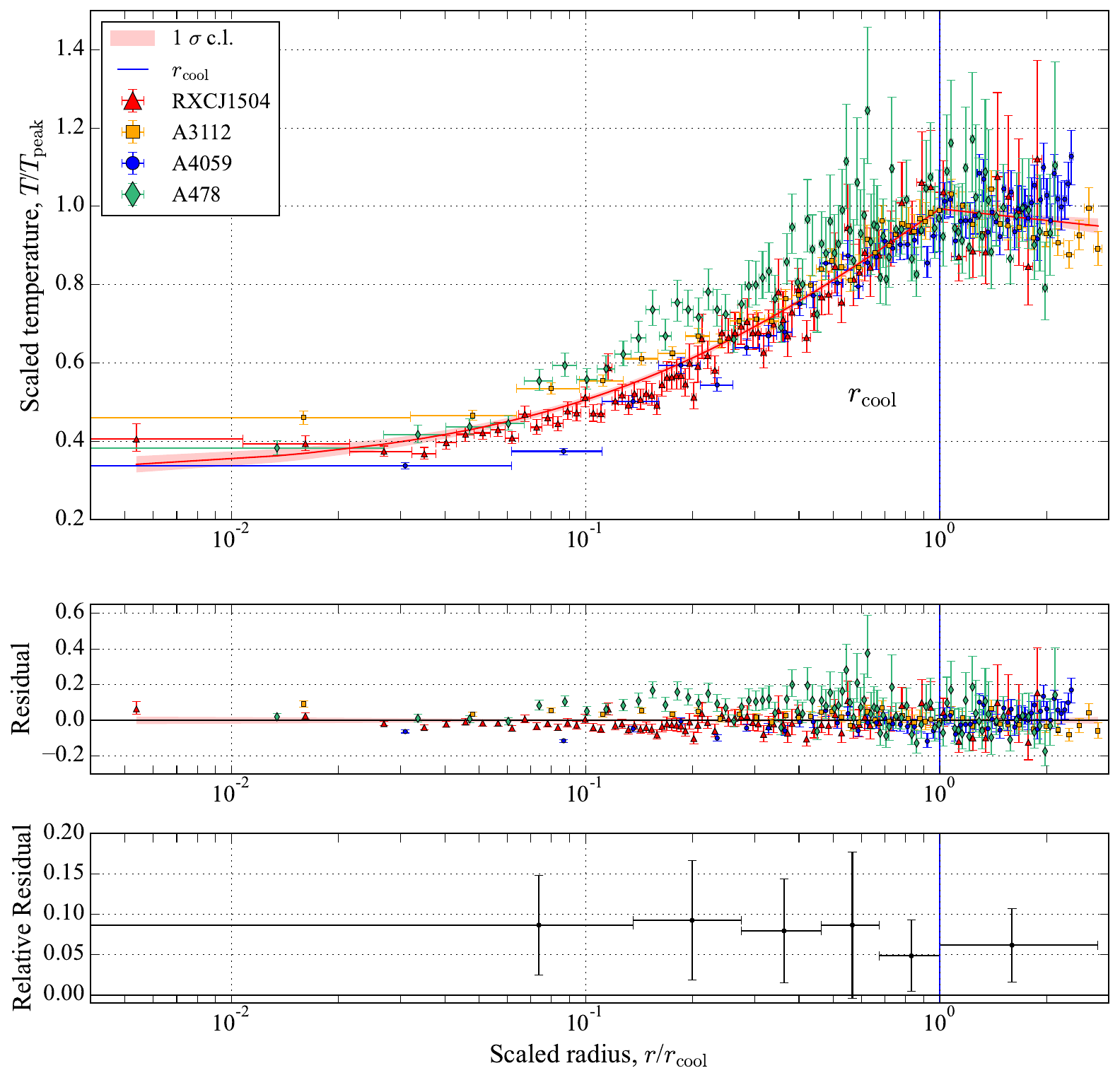}
    \end{center}
   \caption{  
Temperature profile of each cluster scaled by $r_{\rm cool}$ and $T_{\rm peak}$ for the horizontal and the vertical axes, respectively. The red dashed line and shaded region display the best-fit profile and 1\,$\sigma$ uncertainty, respectively. The blue, vertical solid line indicates $r_{\rm cool} = 1$. The residuals and mean absolute relative residuals are shown in the middle and bottom rows, respectively.
}
   \label{fig:norkT_fitting}
   \end{figure}

\begin{figure*}
    \begin{center}
     \includegraphics[width=8.5cm]{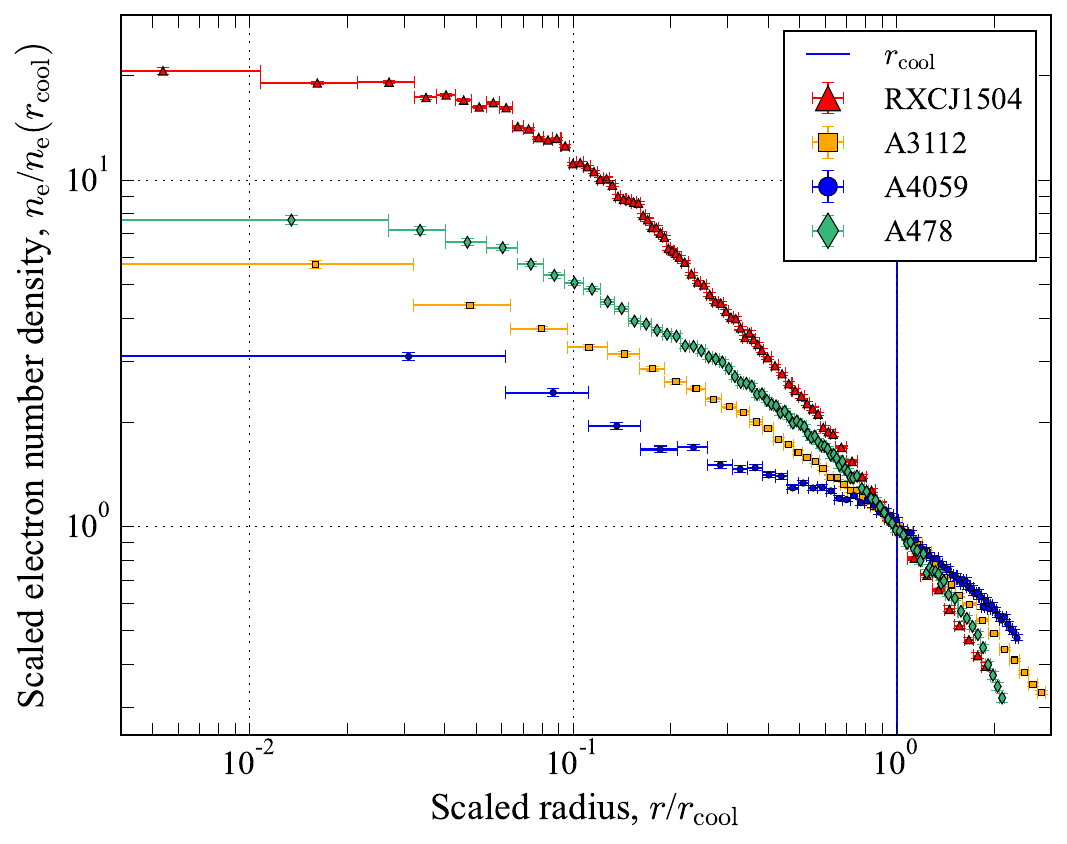}
     \includegraphics[width=8.5cm]{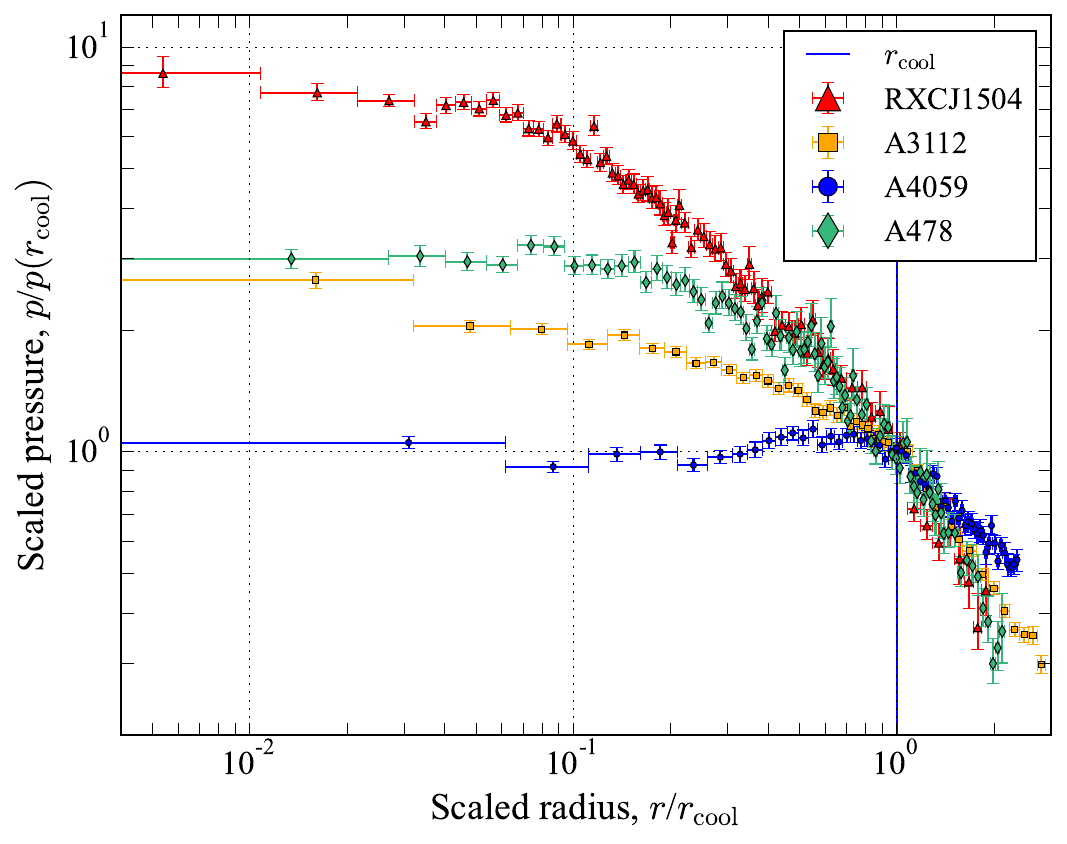}
     \includegraphics[width=8.5cm]{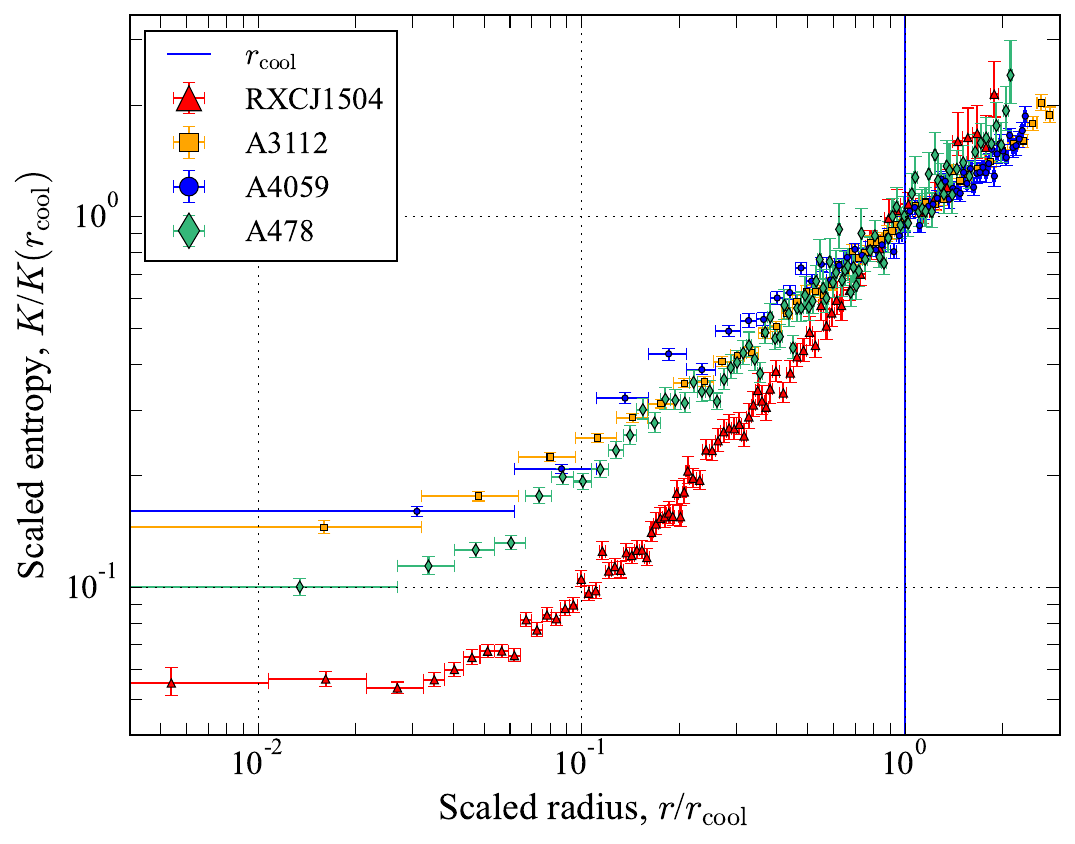}
     \includegraphics[width=8.5cm]{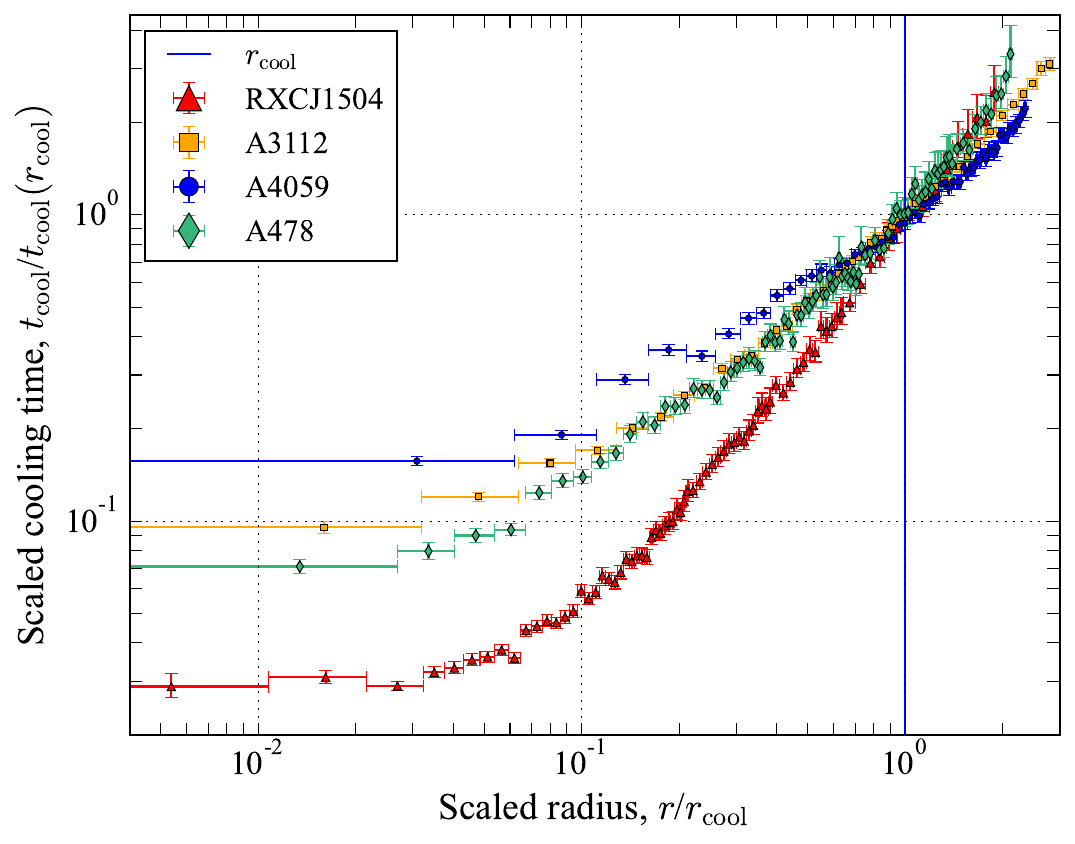}
    \end{center}
   \caption{
Scaled radial profiles of the ICM electron number density (top left), pressure (top right), entropy (bottom left), and radiative cooling time (bottom right) in the sample. The radial profiles are normalized in the same manner as in Figure~\ref{tab:nor_best}.
}
\label{fig:nor_plot}
\end{figure*}

To look deeply into the ICM thermodynamic properties within the cool cores, we, here, scale the observed radial profiles. These scaled profiles are presented by normalizing the values on the horizontal and the vertical axes by the cool-core radius and the value of each component at the cool-core radius, respectively. Figure~\ref{fig:norkT_fitting} shows the scaled radial profile of the ICM temperature. Figure~\ref{fig:nor_plot} presents the scaled profiles of the other components.

A possible universal form is found in the scaled temperature profiles. To clarify this point, we simultaneously fit the scaled temperature profiles using the same model as that used in Section~\ref{sec:temp_fitting} with the MCMC method. In this fitting, we fix $r_{\rm cool}$ in Equation~\ref{eqt:kT} at 1.0 since the temperature profile has already been scaled by $r_{\rm cool}$. Instead, we add a new parameter, intrinsic scatter, into the log-likelihood function. Therefore, the log-likelihood function can be expressed as
\begin{equation}
-2 \ln \mathcal{L} = \sum_{i} \ln{(2 \pi \sigma_{i}^2)} + \sum_{i} \frac{[y_{i} - T(r_{i})]^{2}}{\sigma_{i}^2},
\end{equation}
where $i$ runs over all annulus in which sample, $y_{i}$ and $T(r_i)$ are the scaled value and the model prediction of the scaled temperature profile in each radial bin, respectively, and $\sigma_{i}$ includes the observational uncertainty $\sigma_{y_i}$ and lognormal intrinsic scatter $\sigma_{\rm int}$,
\begin{equation}
    \sigma_{i}^2 = \sigma_{y_{i}}^2 + \sigma_{\rm int}^2.
\end{equation}
The $\sigma_{\rm int}$ parameter accounts for the intrinsic scatter around the mean relation due to unaccounted errors and/or astrophysics.
We continue to use uninformative uniform priors on $T_{\rm center}$, $T_{\rm peak}$, $\alpha_1$, $\alpha_2$ and $\ln{\sigma_{\rm int}}$ as $T_{\rm center} \in (0, 2)$, $T_{\rm peak} \in (0, 2)$, $\alpha_1 \in (0, 2)$, $\alpha_2 \in (-2, 0)$ and $\ln{\sigma_{\rm int}} \in (-20, 1)$. We sample the posterior probability distributions of the parameters over the full parameter space allowed by the priors. The best-fit parameters are summarized in Table~\ref{tab:nor_best}.

%%%%%%%%%%%%%%%%%%%%%%%%%%%%
%%%%%%%%%%%%%%%%%%%%%%%%%%%%
%%%%%%%%%%%%%%%%%%%%%%%%%%%%
\begin{table*}[htbp]
    \begin{center}
    \caption{
    Best-fit parameters derived from the simultaneous fitting of the scaled temperature profiles.
    }\label{tab:nor_best}
    \begin{tabular}{cccccc}
    \hline\hline	
    $T_{\rm center}$      & $T_{\rm peak}$        & $r_{\rm cool}$        & $\alpha_1$        & $\alpha_2$        & $\sigma_{\mathrm{int}}$(\%)\\ \hline
    $0.34 \pm 0.13 $ &$0.99 \pm 0.03 $ & 1 (fixed)  &$0.78 \pm 0.12 $ &$-0.14 \pm 0.10 $ &$3.25_{-0.79}^{+1.04}$\\\hline
        \end{tabular}
    \end{center}
\end{table*}
%%%%%%%%%%%%%%%%%%%%%%%%%%%%

We find that the ratio of the temperature at the center to the peak value is $0.34 \pm 0.13$, which is in good agreement with the well-known observational trend in the temperature profiles of cool-core clusters \citep[e.g.,][]{Sanderson2006, Hudson2010, Simionescu2011}. The slopes of the model for the regions inside and outside the cool cores are measured at $\alpha_1 = 0.78 \pm 0.12$ and $\alpha_2 = -0.14 \pm 0.10$, respectively. The intrinsic scatter $\sigma_{\rm int}$ is obtained at $3.25 ^{+1.04}_{-0.79}\,\%$. 

A possible universal form in the temperature profiles scaled by a fiducial radius ($r_{2500}$ or $r_{500}$) within cool cores has been proposed \citep[e.g.,][]{Allen2001, Sanderson2006}. On the other hand, \cite{Vikhlinin2005} and \cite{Hudson2010} reported no such universal form. Our results indicate a possible universal form in the scaled temperature profile. Since all cool cores in the sample are classified into strong cool cores \citep{Hudson2010}, such a universal form may be observed in strong cool cores only. Additionally, we have scaled the temperature profiles by the cool-core radius, rather than a fiducial radius like $r_{500}$. It is possible that the cool-core radius is a more appropriate factor than a fiducial radius for scaling the radial profiles of the ICM thermodynamic properties, as indicated by the observed intrinsic scatter. The ratio of $r_{\rm cool}$ to $r_{500}$ varies within our sample (see Table~\ref{tab:mass}), emphasizing the significance of the chosen radius for scaling in revealing a universal form in the temperature profile.

To investigate possible universal forms seen in the other components, we also scale the radial profiles using the same manner as that for the scaled temperature profiles. Figure~\ref{fig:nor_plot} shows the scaled radial profiles of the ICM electron number density, pressure, entropy, and radiative cooling time. In contrast to the scaled temperature profiles, we find that there is no universal form in the scaled profiles within the cool cores. The values of each profile at the center are highly scattered, suggesting that the ICM temperature is likely the most fundamental factor for characterizing cool cores. However, further studies are required to reveal possible universal forms in the radial profiles of the ICM thermodynamic properties. An approach involving forward model fitting will be useful to conduct a simultaneous analysis of the ICM temperature and density profiles \cite[e.g.,][]{Umetsu2022}.

In the region outside the cool cores, a universal form is found in the scaled entropy profile. Such a universal form is known as the universal entropy profile \citep{Voit2005}. However, in the region inside the cool cores, the scaled entropy profiles for our sample start varying toward the cluster center, indicating that the cool-core radius is influenced by the interplay between cooling and heating processes.

%%%%%%%%%%%%%%%%%%%%%%%%%%%%
%%%%%%%%%%%%%%%%%%%%%%%%%%%%

\subsection{Analysis of thermodynamic perturbations}

It has been studied that thermodynamic perturbations in the ICM are a good proxy for examining gas motions including turbulence \citep[e.g.,][]{Gaspari2014, Churazov2016, Hofmann2016, Ueda2018, Zhuravleva2018, Kitayama2020, Ueda2021, Zhuravleva2023}. \cite{Gaspari2014} presented that entropy perturbations in the ICM can be used to infer one-dimensional Mach numbers of turbulence ($\cal M_{\rm 1D}$), assuming that pressure perturbations in the ICM are negligible. Following \cite{Gaspari2014}, \cite{Hofmann2016} measured pressure and entropy perturbations in the ICM, and estimated $\cal M_{\rm 1D}$ within the cool cores in a sample of 33 galaxy clusters. 

Motivated by \cite{Gaspari2014} and \cite{Hofmann2016}, we investigate thermodynamic perturbations in the ICM to constrain gas motions in the cool cores in our sample. We have measured the residuals and mean absolute relative residuals between the observed and the best-fit profiles (see the middle and bottom panels of Figures~\ref{fig:kT_fitting}, \ref{fig:nD_fitting}, \ref{fig:P_fitting}, and \ref{fig:K_fitting}), which can be converted into average fractional perturbations. We calculate the average fractional perturbations in the ICM thermodynamic properties within the cool cores, as summarized in Table~\ref{tab:dev}. Our results are consistent with those measured by \cite{Hofmann2016} for their sample. In particular, the average fractional perturbations in the ICM pressure in our sample is lower than 0.15, which agrees with their measurements ($0.09 \pm 0.06$). Thus, the observed perturbations are nearly isobaric. Therefore, similar to \cite{Hofmann2016}, the entropy perturbations in the ICM can be used to infer $\cal M_{\rm 1D}$.

Assuming that the perturbations are isobaric, the observed entropy perturbations can directly be converted into values of $\cal M_{\rm 1D}$. A478 exhibits a slightly larger $\cal M_{\rm 1D}$ with $0.077 \pm 0.069$, while the inferred values of $\cal M_{\rm 1D}$ are comparable in the sample. Thus, the corresponding three dimensional Mach numbers ($\cal M_{\rm 3D}$) of the sample is lower than  ${\cal M}_{\rm 3D} < 0.25$, which is significantly lower than unity and is consistent with those measured by the previous studies \citep[][]{Hofmann2016, Hitomi2018, Ueda2021}. We expect that \textit{XRISM} will be able to achieve direct measurements of the turbulent velocity for our sample \citep{Tashiro2018}. Additionally, the Athena X-ray Observatory will provide us with a great opportunity to measure turbulent velocities in cool cores \citep{Nandra2013, Barcons2017}.

%%%%%%%%%%%%%%%%%%%%%%%%%%%%
\begin{table*}[htbp]
    \begin{center}
    \caption{
    Average fractional perturbations in the temperature, electron number density, pressure, and entropy profiles within the cool cores.
    }\label{tab:dev}
    \begin{tabular}{lcccc}
    \hline\hline	
    Cluster         	& $\langle |\mathrm{d}T| /T\rangle$    & $\langle |\mathrm{d}n|/n \rangle$    & $\langle |\mathrm{d}P|/P \rangle$    & $\langle |\mathrm{d}K|/K \rangle$        \\\hline
    RXCJ1504.1-0248		& $0.047 \pm 0.041 $ &$ 0.019 \pm 0.015 $ &$ 0.046 \pm 0.044 $ &$ 0.051 \pm 0.044 $\\
    A3112	        	& $0.021 \pm 0.017 $ &$ 0.029 \pm 0.051 $ &$ 0.032 \pm 0.060 $ &$ 0.030 \pm 0.030 $\\
    A4059	        	& $0.046 \pm 0.035 $ &$ 0.028 \pm 0.019 $ &$ 0.031 \pm 0.027 $ &$ 0.060 \pm 0.043 $\\
    A478		    	& $0.075 \pm 0.066 $ &$ 0.015 \pm 0.010 $ &$ 0.073 \pm 0.060 $ &$ 0.077 \pm 0.069 $\\ \hline
    %Simultaneously fitting & $2.27 \pm 0.51 $ &$ 7.85 \pm 0.59 $\\\hline
    \end{tabular}
    \end{center}
    \end{table*}
%%%%%%%%%%%%%%%%%%%%%%%%%%%%

\section{Summary and Conclusions}
\label{sec:summary}

In this paper, we have conducted a detailed study of the radial profiles of the ICM thermodynamic properties in cool-core systems in a sample of four galaxy clusters (\rxcj, A3112, A4059, and A478) using archival X-ray data from the \Chandra ~X-ray Observatory. The goal of this study was to observe the characteristics of the cool cores in the sample and explore mechanisms to generate the observed characteristics. To this end, we have measured the turnover radius in the radial profile of the ICM temperature and defined the cool-core radius as the turnover radius. We have also studied the thermodynamic properties of the ICM within the cool-core radius and the relation between the cool-core radius and the cluster mass. The main conclusions of this paper are summarized as follows:

\begin{enumerate}

\item Since cool cores are characterized by a significant drop in the ICM temperature toward the cluster center, we defined the cool-core radius as the turnover radius in the ICM temperature profile, allowing us to study the ICM thermodynamic properties in the regions inside and outside the cool cores. We found no apparent feature at the cool-core radius in the radial profiles of the ICM electron density, pressure, entropy, and radiative cooling time. These results indicate that the boundary between inside and outside cool cores is primarily identified in the temperature profile, suggesting that the ICM temperature is the most fundamental factor for characterizing cool cores.

\item In our sample, the radiative cooling time of the ICM at the cool-core radius exceeds 10\,Gyr, with RXCJ1504.1-0248 exhibiting a radiative cooling time of $32^{+5}_{-11}$\,Gyr at its cool-core radius. Such a long time scale indicates that not only radiative cooling but also additional mechanisms may be required to explain the observed properties. Gas sloshing is possible to displace cool gas generated by radiative cooling away from the center and induce mixing between such cool gas and the ambient hot gas in the outer region, leading to a temperature drop. 

\item Based on the analysis with a very limited sample size, we found that the best-fit relation between the cool-core radius and the cluster mass $M_{500}$ in our sample is consistent with a linear relation, which may deviate from other expected relations. Our findings suggest that cool cores are linked to the evolution of galaxy clusters. Cool cores may coevolve with their host galaxy clusters. Since the cluster mass increases owing to mergers, which in turn induce gas sloshing, it is plausible that gas sloshing plays a significant role in the evolution of cool cores. 

\item A possible universal form in the temperature profiles scaled by the cool core radius is found in the cool cores in our sample. Such a universal form has been presented in scaled temperature profiles by a fiducial radius such as $r_{500}$. Our result is in broadly agreement with the previous findings. However, the scaled profiles of the other components within the cool cores are highly scattered, indicating that there is no universal form.

\item The one-dimensional Mach numbers of turbulence ($\cal M_{\rm 1D}$) in the cool cores in the sample are constrained by analyzing the entropy perturbations in the ICM. The inferred $\cal M_{\rm 3D}$ is significantly lower than unity, suggesting that subsonic gas motions are dominant in the cool cores.

\end{enumerate}

\begin{acknowledgments}
We are grateful to the anonymous referee for helpful and constructive comments.
We thank Keiichi Umetsu for fruitful discussions. We also thank H.-Y. Karen Yang for helpful comments.
The scientific results of this paper are based in part on data obtained from the Chandra Data Archive: \dataset[DOI:10.25574/cdc.223]{https://doi.org/10.25574/cdc.223}.
%: ObsID 4935, 5793, 17197, 17669, 17670, 2216, 2516, 6972, 7323, 7324, 13135, 897, 5785, 1669, and 6102.
S.U. acknowledges the support from the National Science and Technology Council of Taiwan (NSTC 111-2811- M-007-008 and 111-2112-M-001-026-MY3).
We thank the ASIAA Summer Student Program 2019 for hospitality and providing us with an opportunity to launch the project.
\end{acknowledgments}


\vspace{5mm}
\facilities{CXO}


\software{
astropy \citep{Astropy2013, Astropy2018},
CIAO \citep{Fruscione2006},
XSPEC \citep{Arnaud1996}
}



\appendix

\section{Region selection for X-ray spectral analysis}
\label{sec:apx_reg}

Here, we summarize detailed information regarding the region selection and radial bin sizes of the regions for the X-ray spectral analysis.

%%%%%%%%%%%%%%%%%%%%%%%%%%%%
\begin{table*}[htbp]
    \begin{center}
    \caption{
    Summary of our region selection for the X-ray spectral analysis: region interval, region range, the number of regions, and the maximum size for our region selection.
    }\label{tab:annulus_regions}
    \begin{tabular}{lccccl}
    \hline\hline	
    Cluster				                & Region interval (arcsec)\tablenotemark{a}     & Region range (arcsec)     & Number of regions     & Maximum size (arcsec)\tablenotemark{b}						        \\ 
    \hline
    \multirow{6}{*}{RXCJ1504.1-0248}    & 1	                    & 0--3				& 3                 & \multirow{5}{*}{180} 			                    \\
                                        & 0.5                   & 3--20             & 34                &                                                   \\
                                        & 1                     & 20--36            & 16                &                                                   \\
                                        & 2                     & 36--60            & 12                &                                                   \\
                                        & 5                     & 60--90            & 6                 &                                                   \\
                                        & 10                    & 90--180           & 9                 &                                                   \\
    \hline
    \multirow{3}{*}{A3112}              & 2 	                & 0--60				        & 30                & \multirow{3}{*}{180}					            \\
                                        & 5                     & 60--100                   & 8                 &                                                   \\
                                        & 10                    & 100--180                  & 8                 &                                                   \\
    \hline
    \multirow{4}{*}{A4059}              & 5	                    & 0--5              & 1                 & \multirow{5}{*}{192}                              \\
                                        & 4                     & 5--25             & 5                 &                                                   \\
                                        & 3                     & 25--160           & 45                &                                                   \\
                                        & 4                     & 160--192          & 8                &                                                   \\
    \hline
    \multirow{4}{*}{A478}               & 4	                    & 0--4				& 1                & \multirow{4}{*}{320}				                \\
                                        & 2                     & 4--110            & 53                &                                                   \\
                                        & 4                     & 110--210          & 25                &                                                   \\
                                        & 10                    & 210--320          & 11                &                                                   \\
    \hline
    \end{tabular}
    \end{center}
    \tablenotetext{a}{
    Radial bin size between the inner and outer major axes of each elliptical annulus region.
    }
    \tablenotetext{b}{
    Size at the outermost region in our region selection.
    }
\end{table*}
    %%%%%%%%%%%%%%%%%%%%%%%%%%%%

\section{Radial profiles of the ICM metal abundance}
\label{sec:appendix_profiles}

Here, we show the radial profiles of the ICM metal abundance for the sample. We find that the ICM metal abundance in A3112 starts slightly decreasing at $\sim 15$\,kpc toward the center, A4059 also exhibits a decrease of the ICM metal abundance from $\sim 20$\,kpc to the center, which is consistent with that measured by \cite{Choi2004}. 

%%%%%%%%%%%%%%%%%%%%%%%%%%%%
\begin{figure*}[htbp]
    \begin{center}
     \includegraphics[width=8.5cm]{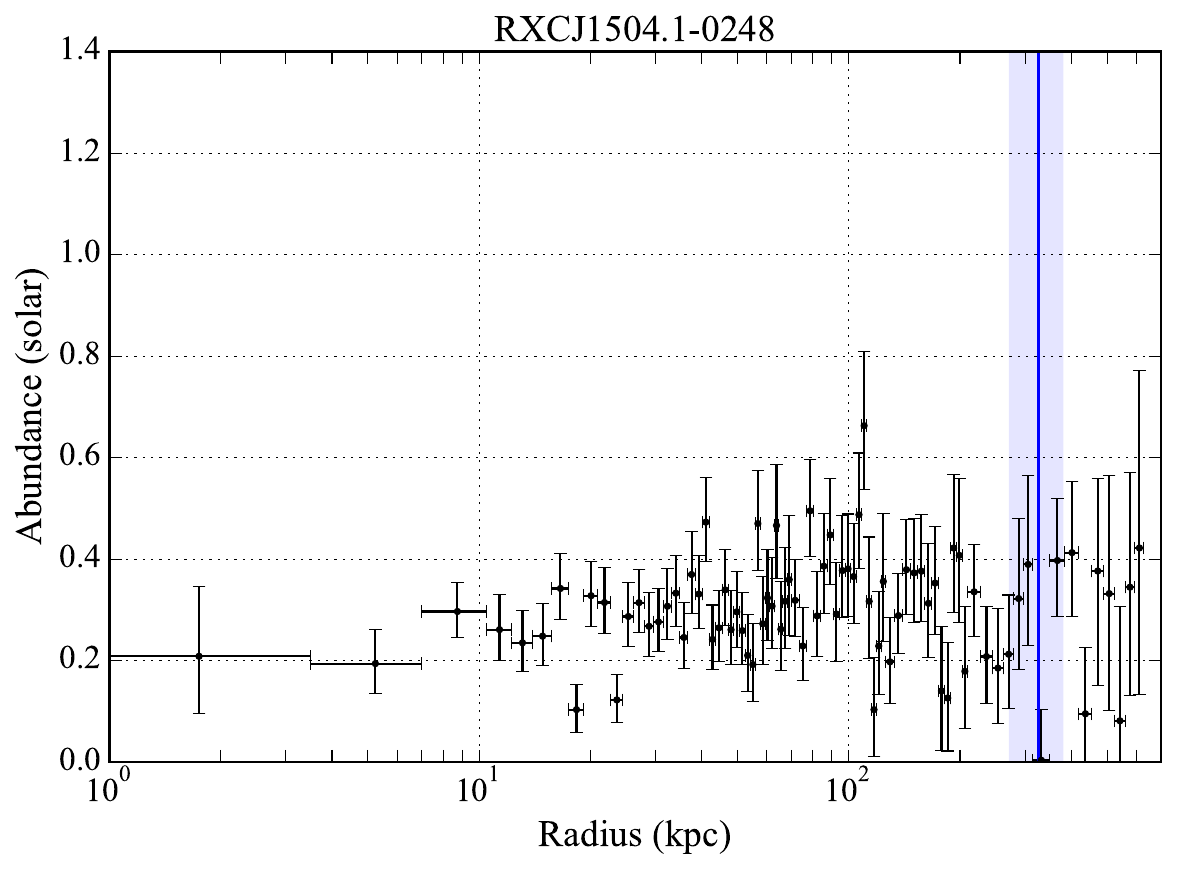}
     \includegraphics[width=8.5cm]{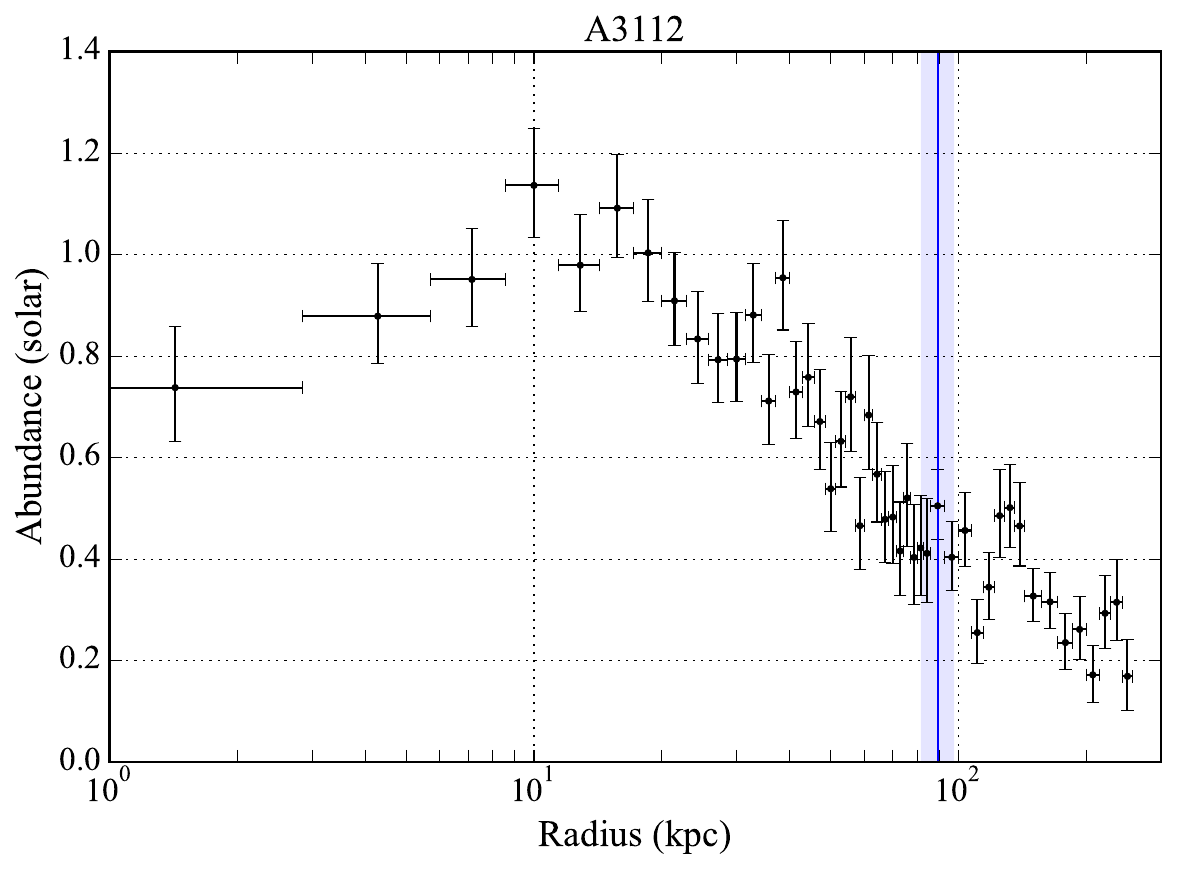}
     \includegraphics[width=8.5cm]{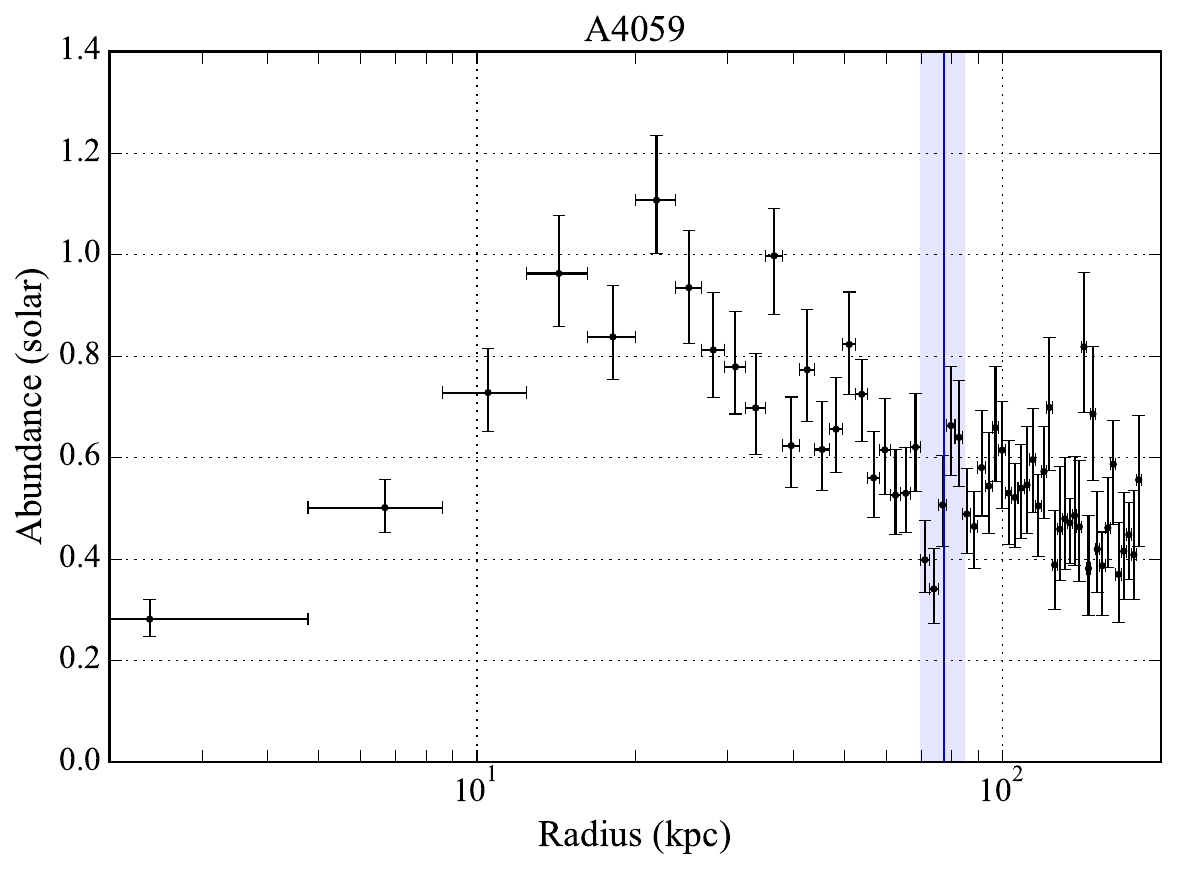}
     \includegraphics[width=8.5cm]{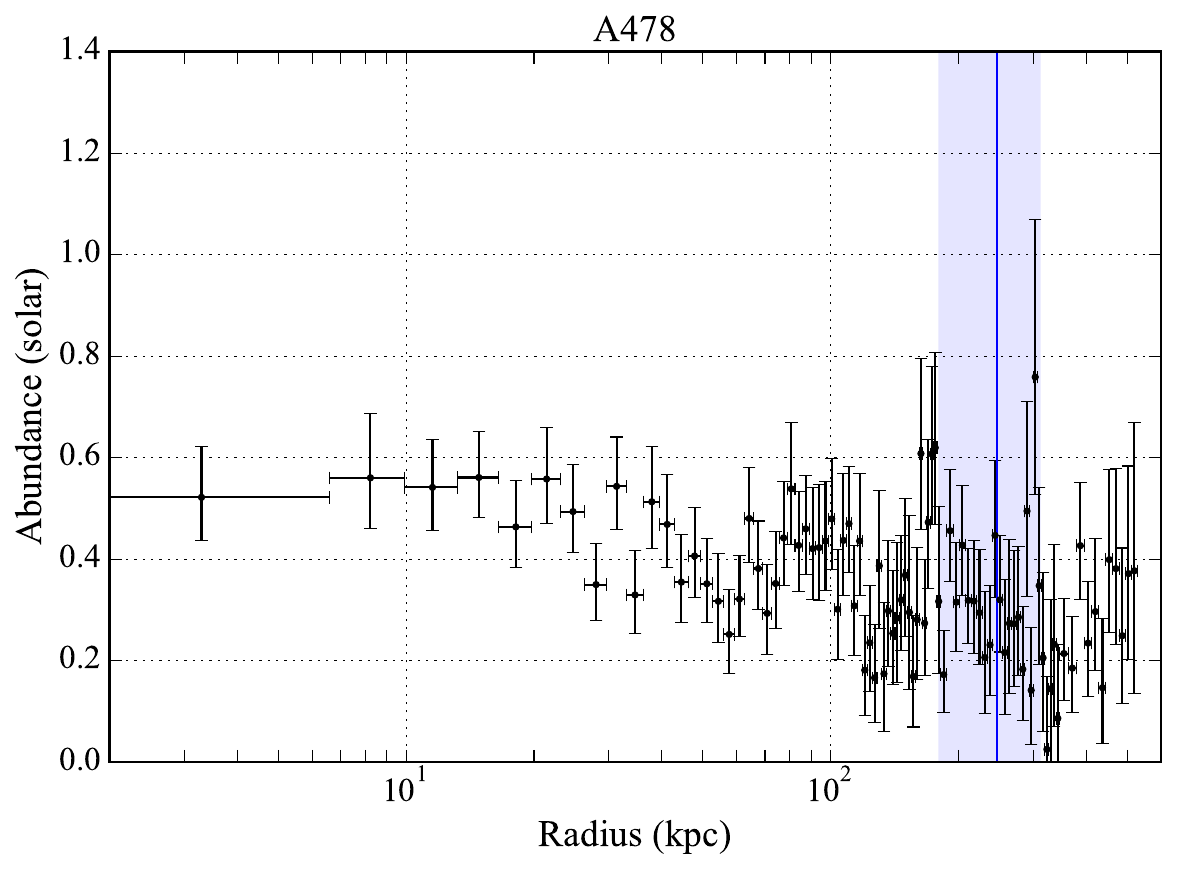}
    \end{center}
   \caption{Radial profiles of the ICM metal abundance: RXCJ1504.1-0248 (top left), A3112 (top right), A4059 (bottom left), and A478 (bottom right). The blue, vertical solid line and shaded region correspond to the cool-core radius and $1\sigma$ uncertainty for each cluster.
      }
\label{fig:abund_plot}
\end{figure*}
%%%%%%%%%%%%%%%%%%%%%%%%%%%%

\clearpage
\bibliography{draft}{}
\bibliographystyle{aasjournal}

\end{document}